\pdfoutput=1

\RequirePackage[l2tabu,orthodox]{nag} 
\documentclass[11pt,letterpaper]{article}

\usepackage[T1]{fontenc}
\usepackage{lmodern}
\usepackage[margin=1.1in]{geometry}
\usepackage{amsmath}
\usepackage{amssymb}
\usepackage{amsthm}
\usepackage[mode=image]{standalone}
\usepackage{array}
\usepackage{enumitem}
\usepackage{url}
\usepackage{booktabs}
\usepackage{siunitx}
\usepackage{graphicx}
\usepackage[
    subrefformat=simple,
    labelformat=simple,
    caption=false,
    font=footnotesize
]{subfig}
\usepackage{cite}
\usepackage[pdfborder={0 0 0}, hidelinks]{hyperref}
\usepackage[capitalize]{cleveref}
\usepackage{textcomp}
\usepackage{microtype}

{}
{}
{}

\DeclareGraphicsExtensions{.pdf}
\graphicspath{{./pdf/}}

\crefformat{equation}{(#2#1#3)}
\crefrangeformat{equation}{(#3#1#4)--(#5#2#6)}
\crefname{figure}{Fig.\@}{Fig.\@}
\Crefname{figure}{Fig.\@}{Fig.\@}

\AtBeginDocument{\DeclareSIUnit{\kWh}{kWh}}
\AtBeginDocument{\DeclareSIUnit{\MWh}{MWh}}
\AtBeginDocument{\DeclareSIUnit{\kva}{kVA}}
\AtBeginDocument{\DeclareSIUnit{\mva}{MVA}}
\AtBeginDocument{\DeclareSIUnit{\pu}{pu}}
\newcommand*\blfootnote[1]{%
  \begingroup
  \renewcommand\thefootnote{}\footnote{#1}%
  \addtocounter{footnote}{-1}%
  \endgroup
}

\DeclareFontFamily{U}{mathx}{\hyphenchar\font45}
\DeclareFontShape{U}{mathx}{m}{n}{<-> mathx10}{}
\DeclareSymbolFont{mathx}{U}{mathx}{m}{n}
\DeclareMathAccent{\widebar}{0}{mathx}{"73}

\begin{document}

\title{Stabilizing Transient Disturbances With Utility-Scale Inverter-Based Resources}

\author{%
    Ryan T. Elliott\protect\\ \texttt{\normalsize ryanelliott@ieee.org}
    \and
    Payman Arabshahi\protect\\ \texttt{\normalsize paymana@uw.edu}
    \and
    Daniel S. Kirschen\protect\\ \texttt{\normalsize kirschen@uw.edu}
}


\date{%
    \vspace*{1ex}
    Department of Electrical and Computer Engineering\protect\\
    University of Washington\protect\\[1\baselineskip]
    November 30th 2020
    \\[1\baselineskip]
    \centering{\textbf{Abstract}}\\[2ex]%
    \normalfont\normalsize%
    \parbox{0.892\linewidth}{%
        This paper presents a trajectory tracking control strategy that
        modulates the active power injected by geographically distributed
        inverter-based resources to support transient stability.  Each
        resource is independently controlled, and its response drives the
        local bus voltage angle toward a trajectory that tracks the angle of
        the center of inertia.  The center-of-inertia angle is estimated in
        real time from wide-area measurements.  The main objectives are to
        stabilize transient disturbances and increase the amount of power that
        can be safely transferred over key transmission paths without loss of
        synchronism.  Here we envision the actuators as utility-scale energy
        storage systems; however, equivalent examples could be developed for
        partially-curtailed photovoltaic generation and/or Type 4 wind turbine
        generators.  The strategy stems from a time-varying linearization of
        the equations of motion for a synchronous machine.  The control action
        produces synchronizing torque in a special reference frame that
        accounts for the motion of the center of inertia.  This drives the
        system states toward the desired trajectory and promotes rotor angle
        stability.  For testing we employ a reduced-order dynamic model of the
        North American Western Interconnection.  The results show that this
        approach improves system reliability and can increase capacity
        utilization on stability-limited transmission corridors.
        \blfootnote{%
            This is a preprint of a paper accepted by IET Generation,
            Transmission, \& Distribution and is subject to Institution of
            Engineering and Technology Copyright. When the final version is
            published, the copy of record will be available at the IET Digital
            Library.
        }
    }
}



\maketitle

\section{Introduction}
When power systems that lack sufficient synchronizing torque are
subjected to a severe disturbance they may fail to maintain rotor
angle stability.  To mitigate this risk, stability limits are imposed
on certain transmission corridors that inhibit the full utilization of
existing thermal capacity.  In turn, this increases the investment and
operation costs of the transmission system.  While power system
stabilizers and turbine governors do respond to large disturbances,
their chief objective is not to maintain first swing transient
stability.  With the exception of voltage regulation, engineers have
generally turned to protection and remedial action schemes to maintain
transient stability rather than feedback control
systems~\cite{anderson:96}.  The challenges of implementing the latter
are twofold.  First, stabilizing transient disturbances requires
actuators that can rapidly respond to events such as faults, line
outages, and generator trips.  Second, as the system is pushed further
away from its original stable equilibrium, constructing a suitable
feedback signal solely from local information becomes increasingly
difficult~\cite{zaborszky:81}.  The combination of wide-area
measurement systems (WAMS) and fast-acting inverter-based resources
(IBRs) enables new approaches to address these problems.

In this paper, we develop and demonstrate a trajectory tracking
control strategy for stabilizing transient disturbances.  It modulates
the active power injected by independently controlled, geographically
distributed IBRs. The controllers are independent in the sense that
they each synthesize their own wide-area reference and calculate their
own control error and corresponding injection.  Here we envision the
actuators as utility-scale energy storage systems (ESS).  The response
of each actuator drives its local bus voltage angle toward a
trajectory that tracks an estimate of the angle of the center of
inertia synthesized from WAMS data.  The overall strategy arises from
a time-varying linearization of the equations of motion for a
synchronous machine.  The control response produces synchronizing
torque in a special reference frame that accounts for the motion of
the center of inertia.  This drives the system states toward the
desired nonequilibrium trajectory and promotes rotor angle stability.
For testing, we employ a reduced-order dynamic model of the North
American Western Interconnection called the
\textit{miniWECC}~\cite{trudnowski:14}.  The results show that this control
strategy improves system reliability and can increase capacity
utilization on stability-limited transmission corridors.

\subsection{Literature Review}
From a systems theory perspective, transient stability pertains to
stability in the sense of Lyapunov.  Hence, it cannot be completely
determined using linear analysis.  One of the oldest and most
important techniques for assessing transient stability is the
\textit{equal area criterion}.  This graphical method provides
valuable intuition about transient disturbances; however, its scope is
limited to two-machine systems and the classical single-machine
infinite bus framework.  To address these limitations, alternative
methods were sought that could be applied to more complex networks of
synchronous machines.  In~\cite{xue:89}, Xue et al.\ developed what
they called the \textit{extended} equal area criterion, which was
based upon a two-machine equivalent model of the system.  In parallel,
there was a movement toward \textit{direct methods} of determining
transient stability, which do not require the solution of the
differential equations, i.e., simulated
trajectories~\cite{varaiya:85,chiang:95}.  Inspired by the efforts of
Magnusson in~\cite{magnusson:47}, energy-based methods for stability
analysis were developed and refined
in~\cite{athay:79,bergen:81,vittal:82,hiskens:89,pai:89}.

Athay et al.\ put forth a seminal application of energy function
analysis in~\cite{athay:79}.  They developed a system-wide energy
function that accounted for the total change in rotor kinetic energy
and potential energy.  While useful, system-wide energy functions did
not always provide insight into the mechanism of instability, or which
critical machines, if any, were susceptible to loss of
synchronism. Motivated by these factors, Vittal developed energy
functions for individual machines using the concept of partial
stability in~\cite{vittal:82}.  The theory of partial energy functions
was further extended in~\cite{michel:83}.  As described in Section
13.7 of \cite{kundur:94}, limitations related to model accuracy and
computational reliability impeded the adoption of direct methods in
practice, but many core ideas were revisited in the context of
\textit{hybrid methods}.  The defining characteristic of hybrid
methods is that they combine energy function analysis with traditional
time-domain simulation~\cite{maria:90,mansour:95,stanton:95}.  This
approach sidesteps some of the challenges associated with direct
methods while streamlining the process of determining stability
margins.  In~\cref{sec:nminus}, we use a hybrid approach inspired
by~\cite{stanton:95} to study the impact of the proposed control
strategy on system faults.  Work continues on the subject of direct
methods, with recent examples including~\cite{vu:16,ju:18}.  For
further recent developments in transient stability assessment,
see~\cite{iravani:20,romay:20,wang:20}.

Numerous protection and remedial action schemes (RAS) have been
developed to bolster transient
stability~\cite{zweigle:13,bhui:17,khorsand:18,yao:19}.  Also called
\textit{system protection schemes}, RAS initiate a predetermined
action, or sequence thereof, in response to a particular condition or
event~\cite{anderson:96}.  In contrast, feedback control systems
modulate the output of one or more actuators in response to an error
signal.  A majority of the control systems developed for transient
stability regulate bus voltages using synchronous machine excitation
systems and/or FACTS devices~\cite{guo:01,li:17}. Related
applications for series devices have also been
explored~\cite{ghandhari:01,haque:08}.  Driven by
economic and environmental factors, many contemporary large-scale
power systems are experiencing rapid growth in the number of
inverter-based resources.  As the penetration of IBRs has grown, so
too has interest in their potential to support transient
stability~\cite{kawabe:12,yousefian:17}.
In this work, we present a control strategy designed to stabilize
transient disturbances by modulating the active power injected by
IBRs.  Examples of suitable actuators include energy storage systems
and partially-curtailed photovoltaic generation.

\subsection{Paper Organization}
The remainder of this paper is organized as follows.
\Cref{sec:method} presents a time-varying linearization of the
equations of motion for a synchronous machine. We then derive a
trajectory tracking control strategy that emerges from this framework.
In \cref{sec:sensitivity_studies}, we examine a set of large-scale
sensitivity studies based on a reduced-order dynamic model of the
Western Interconnection.  \Cref{sec:nminus} outlines the lessons
learned from $\text{N-1}$ contingency analysis.  In
\cref{sec:coi_rating}, we discuss a simplified path rating study for
the California-Oregon Intertie.  \Cref{sec:conclusion} summarizes and
concludes.

\clearpage
\section{Proposed Method}
\label{sec:method}
We base the wide-area control strategy on a time-varying linearization
of the equations of motion for a synchronous machine. The accuracy of
classical linear time-invariant (LTI) models tends to decrease when
the system operating point is driven away from the initial
equilibrium.  Linearizing the system dynamics around an appropriately
selected trajectory can improve model accuracy under transient
disturbances~\cite{elliott:19,elliott:20}.  This section provides a
primer on the definitions and theory of continuous-time linear
time-varying (LTV) systems.  It then presents a derivation showing how
the control strategy arises from these conceptual foundations.

\subsection{Mathematical Preliminaries}
\label{sec:preliminaries}

Let ${f:\mathbb{R}^{n}\times\mathbb{R}^{m}\rightarrow\mathbb{R}^{n}}$
denote a nonlinear vector field
\begin{equation}
    \label{eq:nl_vector_field}
    \dot{x}(t) = f(x(t), u(t)),
\end{equation}
where ${x(t) \in \mathbb{R}^{n}}$ is the system state at time $t$ and
${u(t) \in \mathbb{R}^{m}}$ the input. Linearizing $f$ about a static
equilibrium $\{x_{0},u_{0}\}$ produces a linear time-invariant system
representation.  Alternatively, linearizing about a nonequilibrium
trajectory $\{\widebar{x}(t), \widebar{u}(t)\}$ produces a linear
time-varying representation
\begin{equation}
    \label{eq:ltv_state_space}
    \Delta \dot{x}(t) = A(t)\Delta x(t) + B(t)\Delta u(t),
\end{equation}
where ${\Delta x(t) = x(t) - \widebar{x}(t)}$ and
${\Delta u(t) = u(t) - \widebar{u}(t)}$.  As explained
in~\cite{tsakalis:93}, the state-space matrices in
\cref{eq:ltv_state_space} are functions of time
\begin{align}
    \label{eq:system_matrix}
    A(t) &= D_{x}f(\widebar{x}(t),\widebar{u}(t)) \\
    \label{eq:input_matrix}
    B(t) &= D_{u}f(\widebar{x}(t),\widebar{u}(t)),
\end{align}
where the operator $D_{x}$ returns the Jacobian matrix of partial
derivatives with respect to $x$ evaluated at time $t$, and $D_{u}$
returns the analogous matrix of partials with respect to
$u$.

\subsection{Control Strategy Derivation}
\label{sec:derivation}

In this section, we derive a trajectory tracking control strategy by
applying the concepts introduced in \cref{sec:preliminaries}.
\Cref{tab:notation} provides an overview of the mathematical notation
used throughout the derivation.
\begin{table}[!t]
    \renewcommand{\arraystretch}{1.1}
    \centering
    \caption{Mathematical Notation\strut}
    \label{tab:notation}

    \begin{tabular}{l l l}
        \toprule
        Symbol & Meaning & Units \\
        \midrule
        $\delta(t)$ & rotor angle & elect.\ \si{\radian} \\
        $\widetilde{\delta}(t)$ & center-of-inertia angle & elect.\ \si{\radian} \\
        $\widebar{\delta}_{i}(t)$ & desired angle trajectory & elect.\ \si{\radian} \\
        $\omega(t)$ & rotor speed & \si{\pu} \\
        $\widetilde{\omega}(t)$ & center-of-inertia speed & \si{\pu} \\
        $\widebar{\omega}_{i}(t)$ & desired speed trajectory & \si{\pu} \\
        $\omega_{b}$ & speed base & \si{\radian\per\second} \\
        $\omega_{0}$ & synchronous speed & \si{\pu} \\
        $\theta(t)$ & bus voltage angle & \si{\radian} \\
        $\widetilde{\theta}(t)$ & wide-area angle reference & \si{\radian} \\
        $\widebar{\theta}_{j}(t)$ & desired bus voltage angle trajectory & \si{\radian} \\
        $f(t)$ & bus frequency & \si{\hertz} \\
        $\widetilde{f}(t)$ & center-of-inertia frequency & \si{\hertz} \\
        $\widebar{f}_{j}(t)$ & desired bus frequency trajectory & \si{\hertz} \\
        $P_{m}(t)$ & mechanical power & \si{\pu} \\
        $P_{e}(t)$ & electrical power & \si{\pu} \\
        $H$ & inertia constant & \si{\second} \\
        $D$ & LTI damping coefficient & \si{\pu} \\
        $\mathfrak{D}$ & LTV damping coefficient & \si{\pu} \\
        $\mathfrak{T}$ & LTV synchronizing coefficient & \si{\pu} \\
        $\gamma_{k}$ & sensor weights & -- \\
        $\alpha_{1},\alpha_{2}$ & tuning parameters & -- \\
        $\Delta\xi(t)$ & control error & -- \\
        $i$ & synchronous machine index & -- \\
        $j$ & actuator index & -- \\
        $k$ & sensor index & -- \\
        \bottomrule
    \end{tabular}%
\end{table}
Consider a synchronous machine connected to a large power system.  In
terms of per-unit accelerating power, the nonlinear equations of
motion may be expressed as
\begin{align}
    \label{eq:f1_speed}
    \dot{\delta}(t) &= \omega_{b}\bigl\lbrack\omega(t) - \omega_{0}\bigr\rbrack \\
    \label{eq:f2_accel}
    \dot{\omega}(t) &= -\frac{D}{2H}\bigl\lbrack\omega(t) - \omega_{0}\bigr\rbrack
    + \frac{1}{2H\omega(t)}\bigl\lbrack P_{m}(t) - P_{e}(t)\bigr\rbrack,
\end{align}
where $\omega_{0}$ is the per-unit synchronous speed, and
${\omega_{b}=2\pi f_{0}}$ is the per-unit electrical speed base, where
$f_{0}$ is the nominal system frequency. The machine damping
coefficient is denoted by $D$, and the inertia constant by $H$.
Recall that \cref{eq:f1_speed} describes the angular velocity of the
rotor and \cref{eq:f2_accel} the acceleration.  Now let
$\mathfrak{D}(t)$ and $\mathfrak{T}(t)$ be time-varying coefficients
defined as
\begin{align}
    \label{eq:ltv_coefficients}
    \left.
    \begin{bmatrix}
        \mathfrak{D}(t) \\[0.5ex]
        \mathfrak{T}(t)
    \end{bmatrix} \triangleq
    -2H
    \begin{bmatrix}
        \dfrac{\partial \dot{\omega}}{\partial\omega} \\[2.5ex]
        \dfrac{\partial \dot{\omega}}{\partial\delta}
    \end{bmatrix}\right\rvert_{\widebar{x}(t),\widebar{u}(t)},
\end{align}
where the right-hand side represents the scaled partial derivatives of
\cref{eq:f2_accel} taken with respect to the state variables $\omega$
and $\delta$. These derivatives are evaluated about a state trajectory
$\widebar{x}(t)$ and input $\widebar{u}(t)$.  We show that when
$\widebar{x}(t)$ is selected appropriately, $\mathfrak{D}(t)$ is the
damping coefficient in the center-of-inertia reference frame and
$\mathfrak{T}(t)$ the synchronizing torque coefficient. For analysis
of the LTV damping coefficient $\mathfrak{D}(t)$,
see~\cite{elliott:19}.

The appropriate analytical form of $\mathfrak{T}(t)$ depends on the
model being used to describe transmission network.  In the special
case of a single-machine infinite bus system, the electrical
power output of the machine is given by
\begin{equation}
    \label{eq:classical_p_elect}
    P_{e}(t) = \frac{EV}{X}\sin\delta(t),
\end{equation}
where $X$ is the sum of the synchronous reactance and the line
reactance between the terminals of the machine and the infinite bus.
The internal stator voltage magnitude is denoted by $E$ and the
voltage magnitude at the infinite bus by $V$.  Per convention, the
infinite bus has a voltage angle of zero. Evaluating
\cref{eq:ltv_coefficients}, the LTV synchronizing torque coefficient
is
\begin{equation}
    \label{eq:ltv_synchronizing}
    \mathfrak{T}(t) = \frac{EV}{X}\frac{\cos{\widebar{\delta}(t)}}{\widebar{\omega}(t)},
\end{equation}
where $\widebar{\omega}(t)$ and $\widebar{\delta}(t)$
specify the nonequilibrium state trajectory generically
represented as $\widebar{x}(t)$.
The expression in \cref{eq:ltv_synchronizing} could be readily
modified to accommodate a third-order Heffron-Phillips model by
allowing the internal stator voltage to vary with time.  In the
context of multi-machine models such as those described
in~\cite{varaiya:85}, $\mathfrak{T}(t)$ follows
from~\cref{eq:ltv_coefficients}.


Linearizing \cref{eq:f1_speed} and \cref{eq:f2_accel} about a
trajectory and expressing the result in terms of $\mathfrak{D}(t)$ and
$\mathfrak{T}(t)$ yields
\begin{align}
    \label{eq:ltv_f1_speed}
    \Delta\dot{\delta}(t) ={}& \omega_{b}\Delta\omega(t) \\
    \label{eq:ltv_f2_accel}
    \Delta \dot{\omega}(t) ={}& {-\frac{\mathfrak{D}(t)}{2H}\Delta\omega(t)}
    - \frac{\mathfrak{T}(t)}{2H}\Delta\delta(t)
    + \frac{1}{2H\widebar{\omega}(t)}\Delta P_{m}(t),
\end{align}
where the state deviations are given by
${\Delta{\delta}(t) = \delta(t) - \widebar{\delta}(t)}$ and
${\Delta{\omega}(t) = \omega(t) - \widebar{\omega}(t)}$.  Recall that
$\widebar{\delta}(t)$ denotes the nonequilibrium angle trajectory, and
$\widebar{\omega}(t)$ the speed trajectory.  Furthermore,
${\Delta{P}_{m}(t) = P_{m}(t) - \widebar{P}_{m}(t)}$, where
$\widebar{P}_{m}(t)$ is the mechanical power input trajectory.
As shown in~\cite{elliott:19,elliott:20}, it is possible to increase
damping and support small-signal stability by producing electrical
torque that is in phase with the speed deviation between the rotor and
the center of inertia.  Here we develop a strategy for stabilizing
transient disturbances that produces an electrical power injection
that is in phase with~$\Delta{\delta}(t)$.

\subsection{Nonequilibrium Trajectory}

The nonequilbrium trajectory about which the equations of motion are
linearized is based on a real-time estimate of the angle of the center
of inertia.  The concept of the center of inertia was introduced
in~\cite{tavora:72} to facilitate a decomposition of the system
dynamics.  This decomposition allowed for precise characterization of
the system frequency and synchronous equilibria.  Let $h_{i}$ be the
normalized inertia constant for the $i$th machine
\begin{equation}
    \label{eq:system_inertia}
    h_{i} = \frac{H_{i}}{H_{\scriptscriptstyle{T}}} \quad\,
    H_{\scriptscriptstyle{T}} = \sum_{i\in\mathcal{I}}H_{i},
\end{equation}
where $\mathcal{I}$ is the set of all online synchronous machines.
The center-of-inertia speed and angle are then defined such that
\begin{align}
    \label{eq:coi_speed_def}
    \widetilde{\omega}(t) &= \sum_{i\in\mathcal{I}}h_{i}\omega_{i}(t) \\
    \label{eq:coi_angle_def}
    \widetilde{\delta}(t) &= \sum_{i\in\mathcal{I}}h_{i}\delta_{i}(t).
\end{align}
Let $\{\widebar{\delta}_{i}(t),\widebar{\omega}_{i}(t)\}$ be the
nonequilibrium state trajectory.  In the control strategy,
$\widebar{\delta}_{i}(t)$ represents the desired angle trajectory
toward which unit $i$ is driven.

As explained in \cite{vittal:82}, the transient kinetic energy
responsible for the separation of one or more generators from the rest
of the system originates from the motion of the critical group of
machines away from the center of inertia. Hence, following a
disturbance, driving the speed of each generator back toward the speed
of the center of inertia serves to reduce this transient kinetic
energy and mitigate the risk of separation (see
\cref{sec:nminus}). The nonequilibrium speed trajectory corresponding
to this strategy is
\begin{equation}
    \label{eq:speed_trajectory}
    \widebar{\omega}_{i}(t) = \widetilde{\omega}(t),
\end{equation}
for all $i$ in $\mathcal{I}$.  Because the rotor speed and angle are
dynamically linked, \cref{eq:speed_trajectory} has implications for
the angle trajectory $\widebar{\delta}_{i}(t)$.
From \cref{eq:f1_speed}, the center-of-inertia angle
$\widetilde{\delta}(t)$ and the desired angle trajectory
$\widebar{\delta}_{i}(t)$ may be stated as
\begin{align}
    \label{eq:coi_angle_int}
    \widetilde{\delta}(t) &= \widetilde{\delta}(t_{0})
    + \omega_{b}\int_{t_{0}}^{t}{\widetilde{\omega}(\tau) - \omega_{0}\,d\tau} \\
    \label{eq:unit_angle_int_one}
    \widebar{\delta}_{i}(t) &= \widebar{\delta}_{i}(t_{0})
    + \omega_{b}\int_{t_{0}}^{t}{\widebar{\omega}_{i}(\tau) - \omega_{0}\,d\tau}.
\end{align}
As a result of \cref{eq:speed_trajectory}, we may restate
\cref{eq:unit_angle_int_one} in terms of $\widetilde{\omega}(t)$
\begin{equation}
    \label{eq:unit_angle_int_two}
    \widebar{\delta}_{i}(t) = \widebar{\delta}_{i}(t_{0})
    + \omega_{b}\int_{t_{0}}^{t}{\widetilde{\omega}(\tau) - \omega_{0}\,d\tau}.
\end{equation}
The center-of-inertia speed appears in the integrand of both
\cref{eq:coi_angle_int} and \cref{eq:unit_angle_int_two} because
${\widebar{\omega}_{i}(t) = \widetilde{\omega}(t)}$ for all
$i\in\mathcal{I}$.

Recall that the LTV angle deviation of the $i$th machine is defined as
${\Delta\delta_{i}(t) = \delta_{i}(t)-\widebar{\delta}_{i}(t)}$. In
order for the angle deviation to be zero in steady state, it must hold
that
\begin{equation}
    \label{eq:steady_state_angle}
    \widebar{\delta}_{i}(t_{0}) = \delta_{i}(t_{0}),
\end{equation}
where $t_{0}$ indicates a point in time where the system resides at a
stable equilibrium. As a consequence of \cref{eq:steady_state_angle},
$\widebar{\delta}_{i}(t)$ is, in general, different for each
unit. Combining \cref{eq:coi_angle_int}, \cref{eq:unit_angle_int_two},
and \cref{eq:steady_state_angle} to solve for the nonequilibrium angle
trajectory yields
\begin{equation}
    \label{eq:angle_trajectory}
    \widebar{\delta}_{i}(t) =
    \widetilde{\delta}(t) - \widetilde{\delta}(t_{0}) + \delta_{i}(t_{0}).
\end{equation}
Hence, $\widebar{\delta}_{i}(t)$ is equal to the angle of the center
of inertia plus a fixed offset
${\delta_{i}(t_{0}) - \widetilde{\delta}(t_{0})}$.  This offset
represents the difference between the rotor angle of the $i$th unit
and the angle of the center of inertia at the stable equilibrium
observed at $t_{0}$.  Thus, the nonequilibrium state trajectory
$\{\widebar{\delta}_{i}(t),\widebar{\omega}_{i}(t)\}$ is fully
described by \cref{eq:speed_trajectory} and
\cref{eq:angle_trajectory}.

\subsection{Real-Time Trajectory Estimation}
\label{sec:trajectory_estimation}

In a real-time control application, the center-of-inertia angle
$\widetilde{\delta}(t)$ would need to be estimated using available
measurements. The challenges posed by this problem fall mostly outside
the scope of this paper; however, we will discuss some core concepts
here.  At the time of this writing, rotor angle measurements are not
generally available through wide-area measurement systems.  Thus, it
is not possible to directly calculate \cref{eq:coi_angle_def} onboard
the controller.  If a suitable estimate of the center-of-inertia speed
is available, $\widetilde\delta(t)$ may be approximated as
\begin{equation}
    \label{eq:coi_angle_approx_int}
    \widetilde{\delta}(t) \approx c
    + \omega_{b}\int_{t_{0}}^{t}{\widetilde{\omega}(\tau) - \omega_{0}\,d\tau},
\end{equation}
where $c$ is a constant.  In practice, \cref{eq:coi_angle_approx_int}
would be computed using numerical integration.  In this paper, we
synthesize the center-of-inertia angle estimates by approximating
\cref{eq:coi_angle_def} with a weighted average of bus voltage angle
measurements
\begin{equation}
    \label{eq:coi_angle_approx}
    \widetilde\delta(t) \approx c + \sum_{k\in\mathcal{K}}{\gamma_{k}\theta_{k}(t)}.
\end{equation}
In \cref{eq:coi_angle_approx}, $\theta_{k}(t)$ is the angle
measurement signal reported by the $k$th sensor, $\gamma_{k}$ the
associated weight, and $\mathcal{K}$ the set of available sensors.
The weights $\gamma_{k}$ are nonnegative and sum to one, i.e.,
${1^{T}\gamma = 1}$. These weights do not necessarily
correspond to physical quantities. They may be thought of as
optimization variables in a problem where the objective is to minimize
the norm of the difference between the estimated and actual
center-of-inertia angle signals.  For simplicity, we consider the
arithmetic mean ${\gamma_{k} = 1/\lvert\mathcal{K}\rvert}$ for all
$k$, where $\lvert\mathcal{K}\rvert$ is the number of available
sensors.  The key outcomes of this paper do not depend strongly on
this choice.

In the controller, the value of $c$ in \cref{eq:coi_angle_approx} is
of little importance because the offset in~\cref{eq:angle_trajectory}
ensures that ${\Delta{\delta}_{i}(t_{0})=0}$. Thus, we may henceforth
regard $c$ as a free parameter or set it to zero.  Given nonideal
sensors, the approach in~\cref{eq:coi_angle_approx} may be slightly
more susceptible to measurement bias
than~\cref{eq:coi_angle_approx_int}; however, the dc offset of the
reference signal is not critical in this application, as noted above.
No generator measurements or data are required to implement the
control strategy. All of the required measurements could be made by
commercially available phasor measurement units (PMUs) as described
in~\cite{phadke:83}.  In practice, radian phase angle measurements are
typically mapped onto the interval $(-\pi,\pi]$.  Using this system,
small physical changes may manifest themselves as large numerical
changes when an angle wraps around the discontinuity at $-\pi$.  Thus,
all phase angle measurements must be properly unwrapped when
estimating $\widetilde\delta(t)$ using~\cref{eq:coi_angle_approx}.
Failure to do so could propagate discontinuities associated with angle
wrapping into the control error itself. For more information on
wide-area phase angle calculations, see~\cite{naspi:16}.

\subsection{Control Structure Refinement}
\label{sec:refinement}

Many synchronous machine excitation systems possess the bandwidth
necessary to produce a field current modulation that is in phase with
$\Delta\delta_{i}(t)$; however, doing so may interfere with the
coordination between the automatic voltage regulator and power system
stabilizer.  Hence, we will explore using inverter-based resources as
the control actuators.  In light of this, we restrict the feedback to
bus voltage angle measurements rather than rotor angles. Let
$\theta_{j}(t)$ be the local voltage angle measured at the $j$th
actuator at time~$t$.  The wide-area angle reference
$\widetilde{\theta}(t)$ is
\begin{equation}
    \label{eq:meas_angle_approx}
    \widetilde\theta(t) = \sum_{k\in\mathcal{K}}{\gamma_{k}\theta_{k}(t)},
\end{equation}
where ${\gamma_{k} = 1/\lvert\mathcal{K}\rvert}$, which follows
from~\cref{eq:coi_angle_approx}. We specify the desired angle
trajectory for the $j$th actuator as

\begin{equation}
    \label{eq:meas_angle_trajectory}
    \widebar{\theta}_{j}(t) =
    \widetilde{\theta}(t) - \widetilde{\theta}(t_{0}) + \theta_{j}(t_{0}).
\end{equation}
This equation follows from the same reasoning as
\cref{eq:angle_trajectory}. The difference is that rather than
beginning with the definition of a desired speed trajectory, we begin
by specifying a frequency trajectory
\begin{equation}
    \widebar{f}_{j}(t) = \widetilde{f}(t),
\end{equation}
where $\widetilde{f}(t)$ is the frequency of the center of
inertia. This small modification is required because the IBRs are not
necessarily co-located with synchronous machines.

In the examples given in this paper, $\widetilde\theta(t)$ was
synthesized from measurements reported by \num{36} simulated sensors
distributed throughout the Western Interconnection.  Of these,
\num{19} were co-located with the actuators, leaving \num{17}
additional sensors distributed throughout the system at other
locations. We envision the sensors as PMUs located at high-voltage
substations.  Each sensor was modeled as a first-order time constant
with ${T=\SI{0.02}{\second}}$.  The effects of measurement noise and
nonideal communication are considered in~\cref{sec:nminus}.

Given an appropriate choice of sign, \cref{eq:ltv_f2_accel} suggests
that synchronizing torque can be produced by injecting real power that
is in phase with
${\Delta{\theta}_{j}(t) = \theta_{j}(t) - \widebar{\theta}_{j}(t)}$.
By analyzing the LTI state deviations we can identify a control
structure that is responsive not only to $\Delta{\theta}_{j}(t)$, but
also to the movement of the center-of-inertia angle away from its
predisturbance value.  The LTI state deviations may be decomposed
into two parts where one component captures the difference between the
state and a time-varying reference, and the other the difference
between the reference and its predisturbance value. The first step in
this procedure is to restate the LTI control error
${\theta_{j}(t) - \theta_{j}(t_{0})}$ in a manner that incorporates
the desired angle trajectory $\widebar{\theta}_{j}(t)$
\begin{align}
    \label{eq:lti_error_split_1}
    \theta_{j}(t) - \theta_{j}(t_{0}) &= \theta_{j}(t) - \theta_{j}(t_{0}) +
    \widebar{\theta}_{j}(t) - \widebar{\theta}_{j}(t) \\
    \label{eq:lti_error_split_2}
    \theta_{j}(t) - \theta_{j}(t_{0}) &=
    \Bigl\lbrack\theta_{j}(t) - \widebar{\theta}_{j}(t)\Bigr\rbrack
    + \Bigl\lbrack\widebar{\theta}_{j}(t) - \theta_{j}(t_{0})\Bigr\rbrack.
\end{align}
The first term on the right-hand side of \cref{eq:lti_error_split_2}
corresponds precisely to the LTV angle deviations
$\Delta{\theta}_{j}(t)$, and the second makes it possible for the
controller to respond to the motion of the reference trajectory away
from the predisturbance equilibrium. If these two terms are weighted
equally, the control error matches the LTI angle deviations;
however, we may instead define the control error $\Delta{\xi}(t)$ as a
linear combination of the two components
\begin{equation}
    \label{eq:control_error_1}
    \Delta{\xi}(t) \triangleq
    \alpha_{1}\Bigl\lbrack\theta_{j}(t) - \widebar{\theta}_{j}(t)\Bigr\rbrack
    + \alpha_{2}\Bigl\lbrack\widebar{\theta}_{j}(t) - \theta_{j}(t_{0})\Bigr\rbrack,
\end{equation}
where $\alpha_{1}$ and $\alpha_{2}$ are tuning parameters restricted
to the unit interval.  These parameters provide the ability to tune
the magnitude of the open-loop frequency response in two distinct
frequency bands, one low and one high.  We examine the relative impact
of these parameters in greater detail in~\cref{sec:sensitivity_studies}.

Rearranging \cref{eq:meas_angle_trajectory}, we see that
\begin{equation}
    \label{eq:angle_trajectory_2}
    \widebar{\theta}_{j}(t) - \theta_{j}(t_{0}) =
    \widetilde{\theta}(t) - \widetilde{\theta}(t_{0}).
\end{equation}
By substitution, the control error $\Delta{\xi}(t)$ in
\cref{eq:control_error_1} may then be restated as
\begin{equation}
    \label{eq:control_error_2}
    \Delta{\xi}(t) =
    \alpha_{1}\Bigl\lbrack\theta_{j}(t) - \widebar{\theta}_{j}(t)\Bigr\rbrack
    + \alpha_{2}\Bigl\lbrack\widetilde{\theta}(t) - \widetilde{\theta}(t_{0})\Bigr\rbrack.
\end{equation}
This form makes it clear that the second term represents the deviation
between the center-of-inertia angle and its predisturbance value.  As
a consequence, this component of the control error is theoretically
the same for all controllers. In practice, there may be small
differences arising from variation in the estimates
of~$\widetilde{\theta}(t)$.

\begin{figure}[!t]
    \centering
    \includegraphics[width=0.75\textwidth]{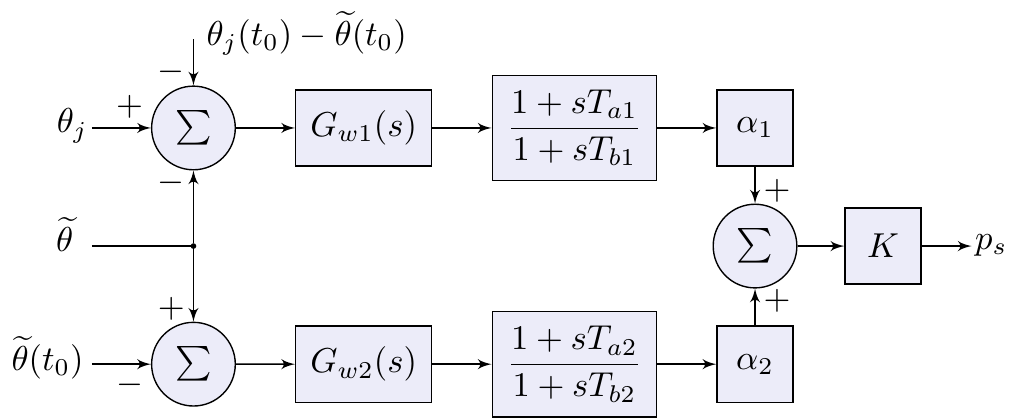}
    \caption{Block diagram of the trajectory tracking wide-area
        synchronizing controller. The $G_{w1}(s)$ and $G_{w2}(s)$
        blocks represent washout (highpass) filters. The lead-lag compensation
        blocks are only used if necessary and may possess any number of
        stages.
    }
    \label{fig:main_control_structure}
\end{figure}

The final structure of the wide-area synchronizing controller is shown
in~\cref{fig:main_control_structure}.  The upper sum at the input
yields the component of the control error multiplied by $\alpha_{1}$,
and the lower sum the component multiplied by $\alpha_{2}$.  There are
two separate compensation paths, each consisting of a washout
(highpass) filter cascaded with a lead-lag compensator.  The form of
the filtering and phase compensation blocks is flexible, and multiple
stages may be employed if necessary.  Tuning considerations
are discussed in~\cref{sec:sensitivity_studies}.  In the case of
energy storage, $p_{s}$ represents a change in the charging setpoint
for positive values of $K$.

\subsection{Converter Interface Description}
\label{sec:converter_interface}

\begin{figure}[!t]
    \centering
    \includegraphics[width=0.65\textwidth]{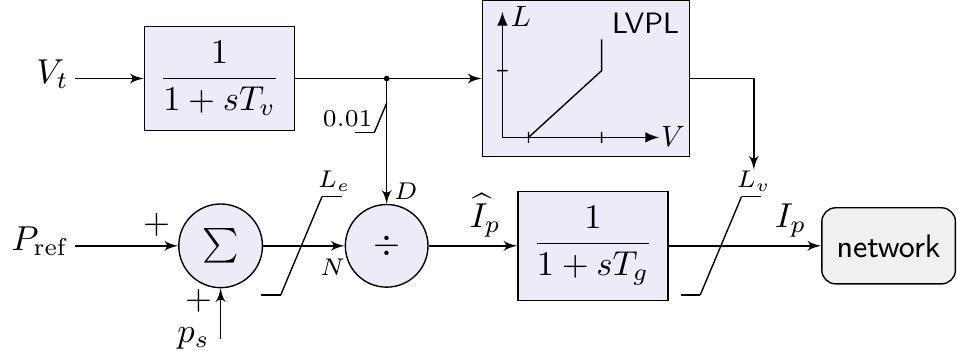}
    \caption{High-level diagram of the simplified converter interface.
        In the division operator $N$ stands for numerator and $D$ for
        denominator. The lower bound on the terminal voltage measurement
        prevents numerical errors and excessively large current commands.
        The \textit{low-voltage power logic} (LVPL) block determines
        voltage-dependent bounds on the injected current.
    }
    \label{fig:converter_interface}
\end{figure}

\Cref{fig:converter_interface} shows a converter interface where for
simplicity it is assumed that the IBR injects only active power.  It
is possible for the IBRs to simultaneously regulate voltage by also
modulating reactive power; however, that functionality has been
omitted to better illustrate the behavior of the proposed control
strategy.  The electrical controller shown
in~\cref{fig:main_control_structure} provides an auxiliary input
$p_{s}$ to this interface. The structure
in~\cref{fig:converter_interface} was based on the generic dynamic
models for renewable energy systems developed in the Western Electric
Coordinating Council (WECC)~\cite{elliott:15}.

Here $P_{\mathrm{ref}}$ denotes the real power setpoint of the IBR,
i.e., the desired baseline injection corresponding to non-control
objectives. The signal $p_{s}$ represents the active power modulation
command computed by the controller in
\cref{fig:main_control_structure}.  In the saturation stage after the
initial sum, the bounds $L_{e}$ on the commanded power account for
device ratings and limits on the ESS state of charge.  The commanded
power is then divided by the terminal voltage magnitude $V_{t}$ to
produce an active current command, $\widehat{I}_{p}$.  In this step, a
lower bound is placed on the voltage measurement to prevent numerical
errors and excessively large current commands.  After accounting for
the interface time constant, the current command is passed through a
final saturation stage where the bounds $L_{v}$ are voltage dependent.
This prevents the model from attempting to inject substantial active
current into a faulted bus.  For more information about this
\textit{low-voltage power logic} (LVPL), see~\cite{elliott:15}. In
simulation, the output of the interface model specifies a boundary
current injection $I_{p}$ for the network equations.

\subsection{Implementation Issues}
\label{sec:implementation}
The primary challenges associated with the implementing the control
strategy developed here revolve around constructing a real-time
estimate of the angle of the center of inertia.  Contemporary PMU
networks are technologically capable of enabling this estimation;
however, logistical challenges remain. For instance, operating
entities may be disinclined to use wide-area measurements recorded
outside of their service territory for real-time
control~\cite{pierre:19}. The reasons for this are mostly related to
policy compliance and cybersecurity~\cite{dolezilek:11}. These issues
are barriers to implementation for wide-area control generally and are
not specific to the strategy developed in this paper. Further
implementation considerations include:
\begin{itemize}
    \item Simulation of full-scale, high-fidelity base cases
    to illuminate any issues potentially missed by reduced-order models;
    \item Analysis of issues such as multi-service provision and
    non-unity power factor operation of the IBRs;
    \item Analysis of the trade-offs associated with estimating the
    center-of-inertia angle locally vs.\ centrally, i.e., at a control
    center;
    \item Design of schemes to ensure the integrity and accuracy
    of the wide-area measurements, including GPS time signals;
    \item Design of supervisory control to ensure that the
    wide-area control strategy is operating as intended~\cite{pierre:16}.
\end{itemize}

\clearpage
\section{Large-Scale Sensitivity Studies}
\label{sec:sensitivity_studies}

\begin{figure}[!b]
    \centering
    \includegraphics[width=0.55\textwidth]{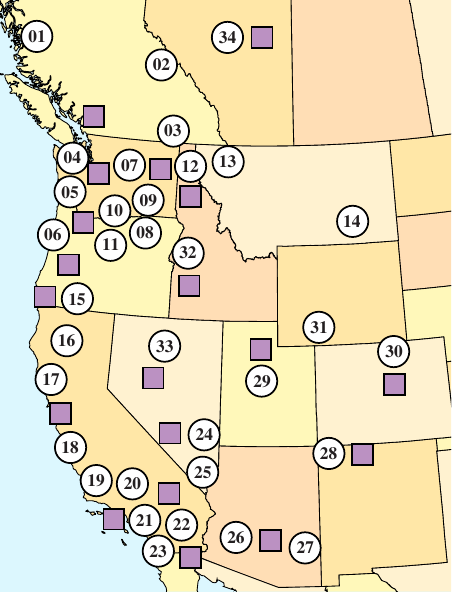}
    \caption{Notable locations in the miniWECC system. The numbered
        circles represent equivalent generators, and the colored squares
        represent the control actuators, which are utility-scale
        inverter-based resources.
    }
    \label{fig:miniwecc_map}
\end{figure}%

To further explore the control strategy introduced
in~\cref{sec:method}, we present a collection of large-scale
sensitivity studies.  For simulation and dynamic analysis, we employ
the MATLAB-based Power System Toolbox (PST)~\cite{chow:92}.  A custom
dynamic model based on the diagrams shown in
Figs.~\ref{fig:main_control_structure} and
\ref{fig:converter_interface} was implemented within the PST
framework.  The test case is based on a reduced-order dynamic model of
the Western Interconnection called the
miniWECC~\cite{trudnowski:14}. The augmented version of the
miniWECC employed here comprises \num{141} buses, \num{190} ac
branches, \num{2} HVDC lines, and \num{34} synchronous generators.  A
total of \num{19} geographically distributed energy storage systems
were installed in the miniWECC, one at each load center.
\Cref{fig:miniwecc_map} shows a map of notable locations in the
system.  The numbered circles indicate equivalent generators, and the
colored squares indicate energy storage systems.  Each generator is
represented using a subtransient reactance model and a compound source
rectifier exciter.  Inputs to the exciter are provided by an automatic
voltage regulator (AVR) and $\Delta\omega$ power system stabilizer
(PSS).  Turbine governors are also represented, taking into account
differences between thermal and hydroelectric units.  In the analysis
performed here, the active component of the system load is modeled as
constant current and the reactive component as constant impedance.

\subsection{Time-Domain Analysis}
\label{sec:time_domain_analysis}

This subsection examines the effect of sweeping the tuning parameters
$\alpha_{1}$ and $\alpha_{2}$ on the time-domain system response.  The
aim is to build intuition about how the control strategy responds to
disturbances, and how the resulting injections affect the state
trajectories.  Each of the~\num{19} participating ESSs, rated
at~\SI{100}{\mega\watt}/\SI{200}{\MWh}, were configured identically
for simplicity.  These systems may correspond either to large
individual ESSs or aggregated storage clusters. We consider a total
ESS power capacity of~\SI{1.9}{\giga\watt}, which represents
about~\SI{1.8}{\percent} of the overall miniWECC load. In all examples
$G_{w1}(s)$ is a first-order highpass filter with a corner frequency
at \SI{0.1}{\hertz}, and $G_{w2}(s)$ second-order with a corner near
\SI{0.01}{\hertz}.  In the $\alpha_{2}$ path, we included
\SI{15}{\degree} of phase lead centered about the \textit{frequency
regulation mode} at \SI{0.02}{\hertz}.  As described
in~\cite{grondin:93,wilches:16}, the frequency regulation mode of a
power system is a very low frequency mode, typically below
\SI{0.1}{\hertz}, in which the rotor speeds of all synchronous
generation units participate.  No phase compensation was introduced in
the $\alpha_{1}$ path.  Lastly, the gain of each controller was held
fixed at ${K = 5}$.

\begin{figure}[!b]
    \centering
    \includegraphics[width=0.631\textwidth]{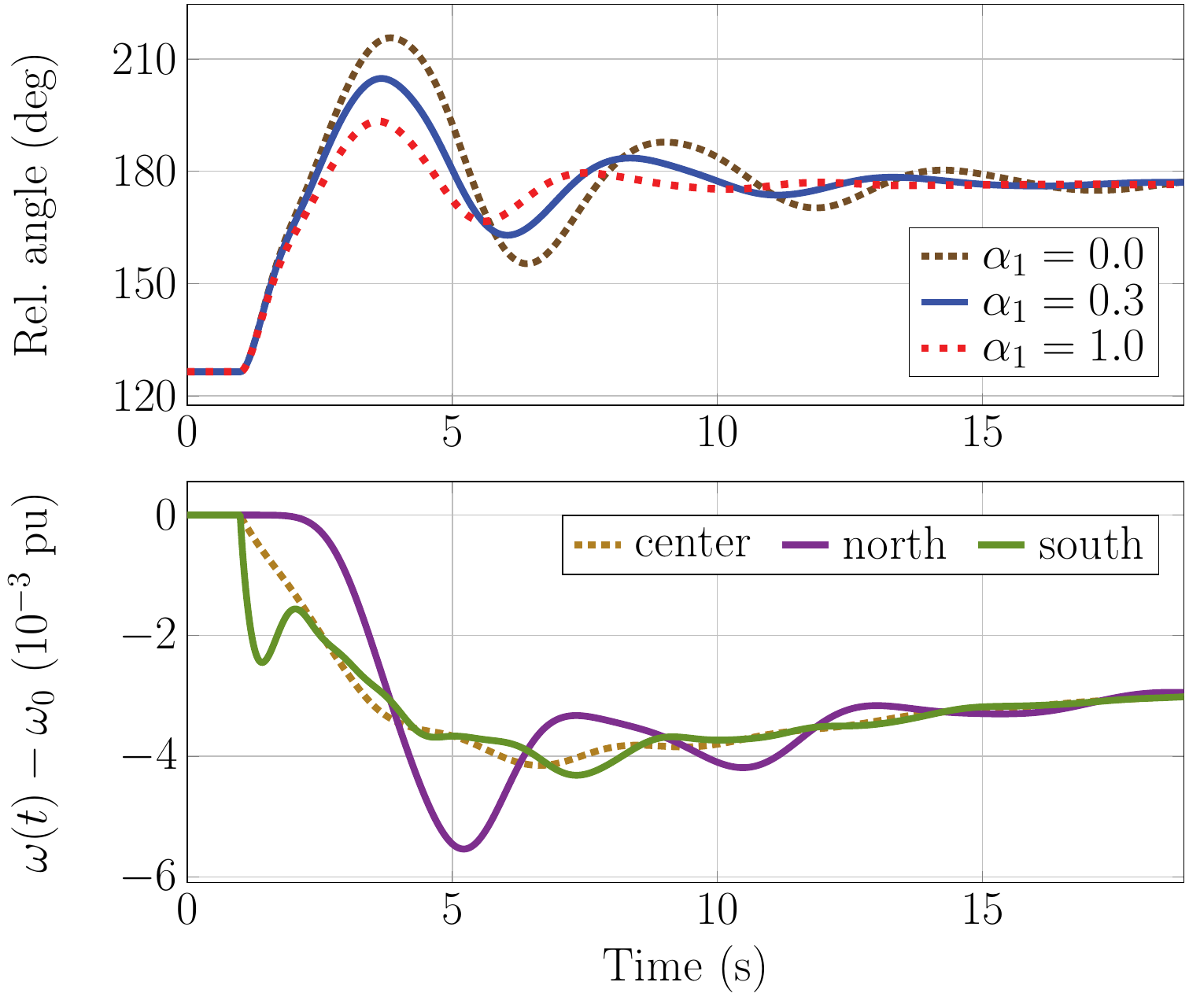}
    \caption{Time-domain simulations of generator G26 in Arizona being
        tripped offline for various values of $\alpha_{1}$, where
        ${\alpha_{2}=0}$.  The top subplot shows the relative angle between
        generator G34 in Alberta (north) and G23 in San Diego (south).  The
        bottom shows the speed of G23, G34, and the center of inertia
        in the open-loop case.
    }
    \label{fig:alpha1_time_sweep_part1}
\end{figure}%

\Cref{fig:alpha1_time_sweep_part1} shows time-domain simulations of a
large nuclear plant in Arizona, G26, being tripped offline for various
values of $\alpha_{1}$, where ${\alpha_{2}=0}$.  This disturbance was
selected because it represents a severe step change in the
generation/demand balance.  The lost generation corresponds to
approximately \SI{4}{\percent} of the total load.  The top subplot
shows the difference between the rotor angle of generator G34 in
Alberta and G23 in San Diego.  This pair represents the maximal
inter-machine angle difference observed for this disturbance.  The
control effort has a stabilizing effect on this angle separation.  In
particular, the frequency of the oscillatory component of the response
rises, which indicates an increase in synchronizing torque.

\begin{figure}[!b]
    \centering
    \includegraphics[width=0.631\textwidth]{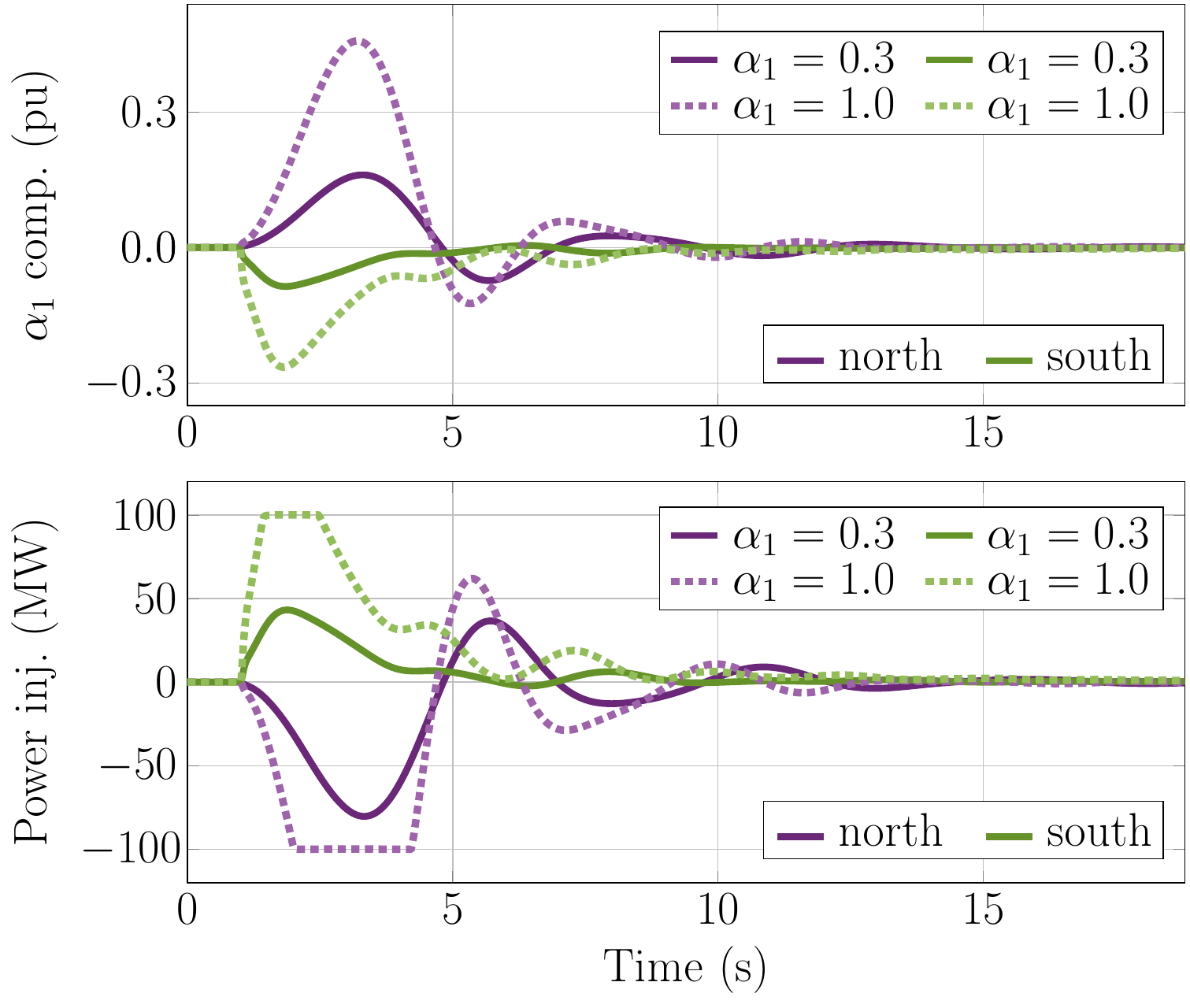}
    \caption{The control response corresponding to the time-domain
        simulations shown in \cref{fig:alpha1_time_sweep_part1}.  The top
        subplot shows the $\alpha_{1}$ component of the control error and the
        bottom the injected power. The purple and green traces show the
        behavior of the actuators located near the load centers in Alberta
        (north) and San Diego (south), respectively.
    }
    \label{fig:alpha1_time_sweep_part2}
\end{figure}%

The bottom subplot of \cref{fig:alpha1_time_sweep_part1} shows the
speed of generator G34 (Alberta) and G23 (San Diego) compared with the
speed of the center of inertia for the open-loop case. Immediately
following the disturbance, the speed of the generator in Alberta is
slightly faster than the speed of the center of inertia, and the speed
of the generator in San Diego slightly slower. This occurs because the
machine in San Diego is electrically closer to the tripped generator
than the average machine in the system, and the machine in Alberta
farther away.  Over the first swing of the transient, these
relationships manifest themselves as a positive $\alpha_{1}$ error
component in Alberta and a negative $\alpha_{1}$ component in San
Diego.  The top subplot of \cref{fig:alpha1_time_sweep_part2} shows
the $\alpha_{1}$ component of the control response, which results from
passing
${\alpha_{1}\lbrack\theta_{j}(t)-\widebar{\theta}_{j}(t)\rbrack}$
through the controller compensation.  To mitigate the system
separation in the first swing, the ESS in Alberta charges and the one
in San Diego discharges. This charging modulation is shown in the
bottom subplot of \cref{fig:alpha1_time_sweep_part2}.

\begin{figure}[!t]
    \centering
    \includegraphics[width=0.631\textwidth]{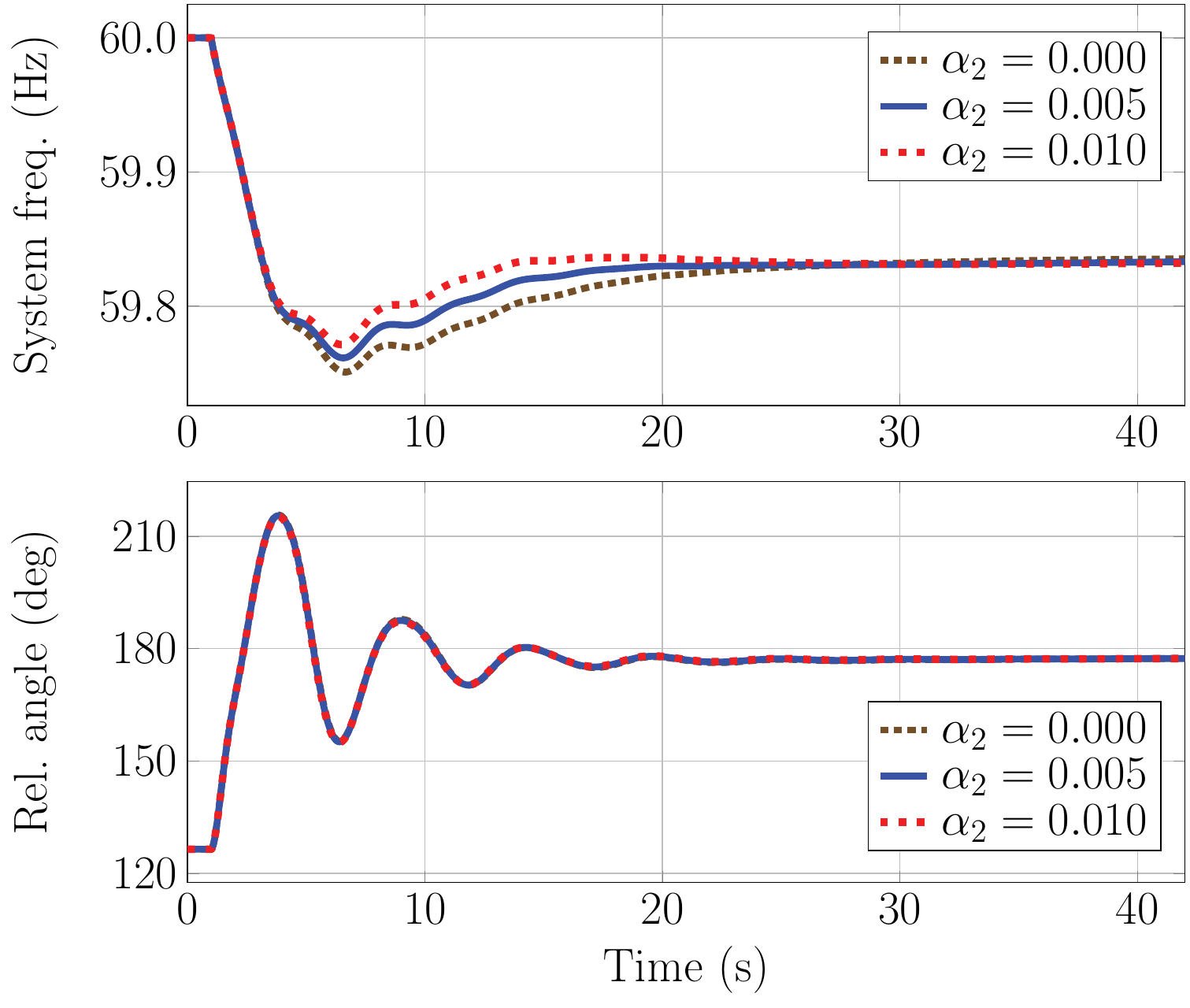}
    \caption{Time-domain simulations of generator G26 in Arizona being
        tripped offline for various values of $\alpha_{2}$, where
        ${\alpha_{1}=0}$.  The top subplot shows the frequency of the center
        of inertia. The bottom shows the relative angle between generator G34
        in Alberta (north) and G23 in San Diego (south). This angle difference
        does not vary with $\alpha_{2}$.
    }
    \label{fig:alpha2_time_sweep_part1}
\end{figure}%

\Cref{fig:alpha2_time_sweep_part1} shows time-domain simulations of
the same generator trip but for the case where $\alpha_{2}$ is swept
over a range of values and ${\alpha_{1}=0}$.  The top subplot shows
the system frequency, calculated as the frequency of the center of
inertia.  The control effort has a stabilizing effect on the frequency
response.  As $\alpha_{2}$ increases the depth of the nadir is
reduced, and the frequency rebounds more quickly. The bottom subplot
shows the relative angle between generator G34 in Alberta and G23 in
San Diego (as in \cref{fig:alpha1_time_sweep_part1}).  By design,
varying the value of $\alpha_{2}$ does not have an impact on this
angle difference. The purpose of $\alpha_{2}$ is solely to adjust the
behavior of the frequency regulation mode, which is common to all
synchronous generators in the system.  In situations where providing
synchronizing torque at the frequency regulation mode is not required
(or desirable), $\alpha_{2}$ may be set to zero with no adverse effect
on the control response over the remainder of the frequency band.

\begin{figure}[!t]
    \centering
    \includegraphics[width=0.631\textwidth]{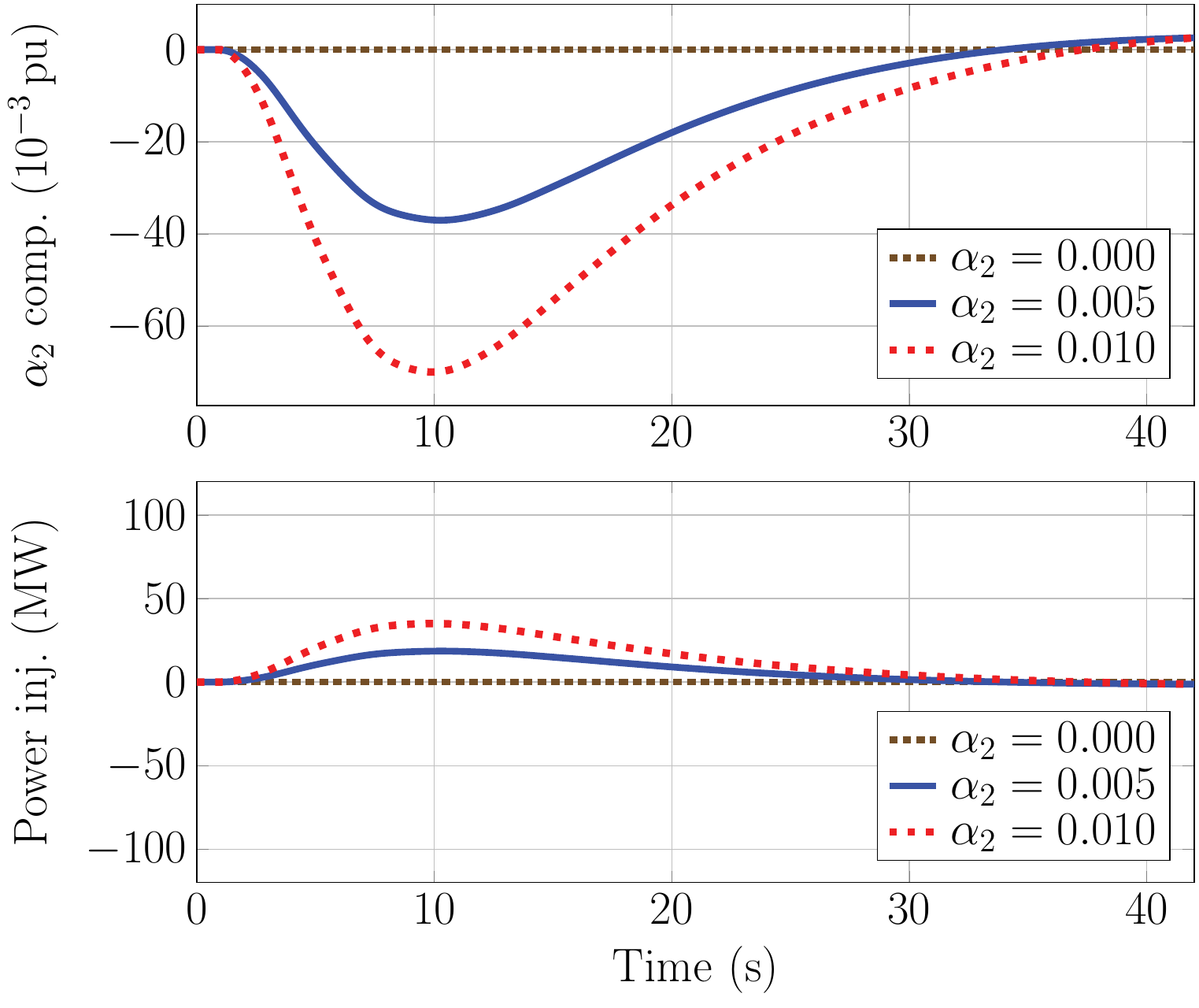}
    \caption{The control response corresponding to the time-domain
        simulations shown in \cref{fig:alpha2_time_sweep_part1}.
        The top subplot shows the $\alpha_{2}$ component of the control
        error and the bottom the injected power.
        In this case all of the controllers synthesize identical
        estimates of $\widetilde{\theta}(t)$, so
        ${\alpha_{2}\lbrack\widetilde{\theta}(t)-\widetilde{\theta}(t_{0})\rbrack}$
        is the same for each actuator.
    }
    \label{fig:alpha2_time_sweep_part2}
\end{figure}%

\Cref{fig:alpha2_time_sweep_part2} shows the behavior of a
representative controller, which is the same for each ESS in this case
(when $\widetilde{\theta}(t)$ is the same).  The top subplot shows the
$\alpha_{2}$ component of the control response, which results from
passing
${\alpha_{2}\lbrack\widetilde{\theta}(t)-\widetilde{\theta}(t_{0})\rbrack}$
through the controller compensation.  Neglecting the effects of
nonideal communication, the center-of-inertia angle
estimate~$\widetilde{\theta}(t)$ is the same for all ESSs resulting in
identical $\alpha_{2}$ components.  (We consider the effects of
nonideal communication in~\cref{sec:nminus}.)  Following the generator
trip, the speed of the center of inertia deflects downward causing
$\widetilde{\theta}(t)$ to decline from its initial value
value~$\widetilde{\theta}(t_{0})$.  This results in a negative
$\alpha_{2}$ component that causes every ESS in the system to inject
power that is in phase with the error.  This is shown in the bottom
subplot of \cref{fig:alpha2_time_sweep_part2}.  In a practical
application, the total control response would effectively be a
superposition of the injections depicted in
Figs.~\ref{fig:alpha1_time_sweep_part2}
and~\ref{fig:alpha2_time_sweep_part2}.

\clearpage
\subsection{Tuning Considerations}
\label{sec:frequency_domain_analysis}

\begin{figure}[!b]
    \captionsetup[subfloat]{farskip=0pt}
    \centering
    \subfloat{
        \centering
        \includegraphics[width=0.385\textwidth]{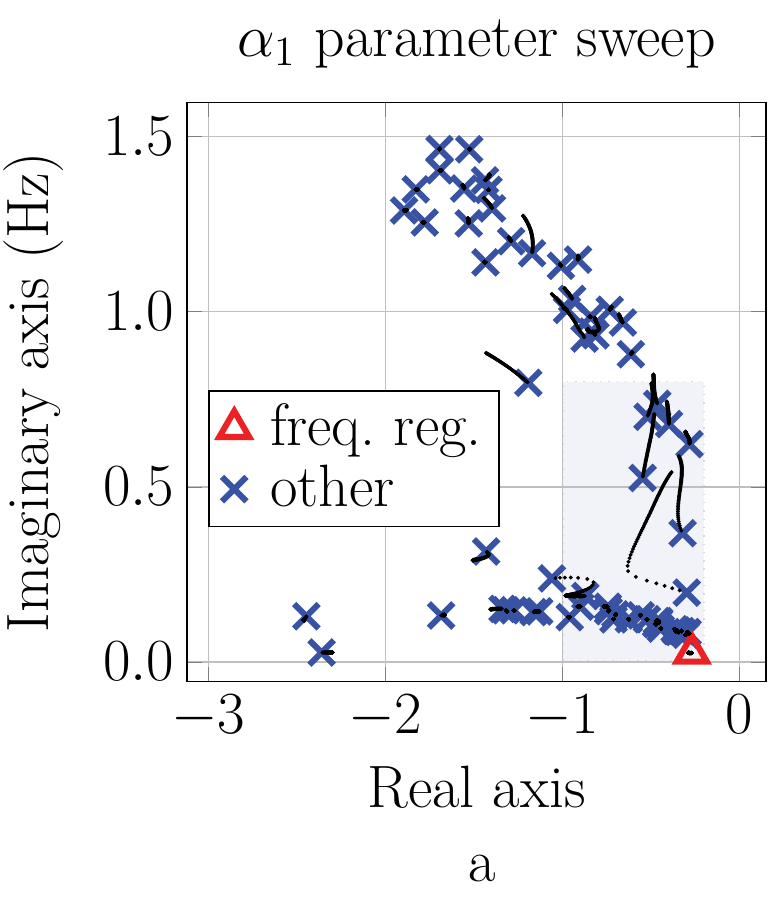}
        \label{fig:mini_alpha1_sensitivity}
    }\qquad
    \subfloat{
        \centering
        \includegraphics[width=0.385\textwidth]{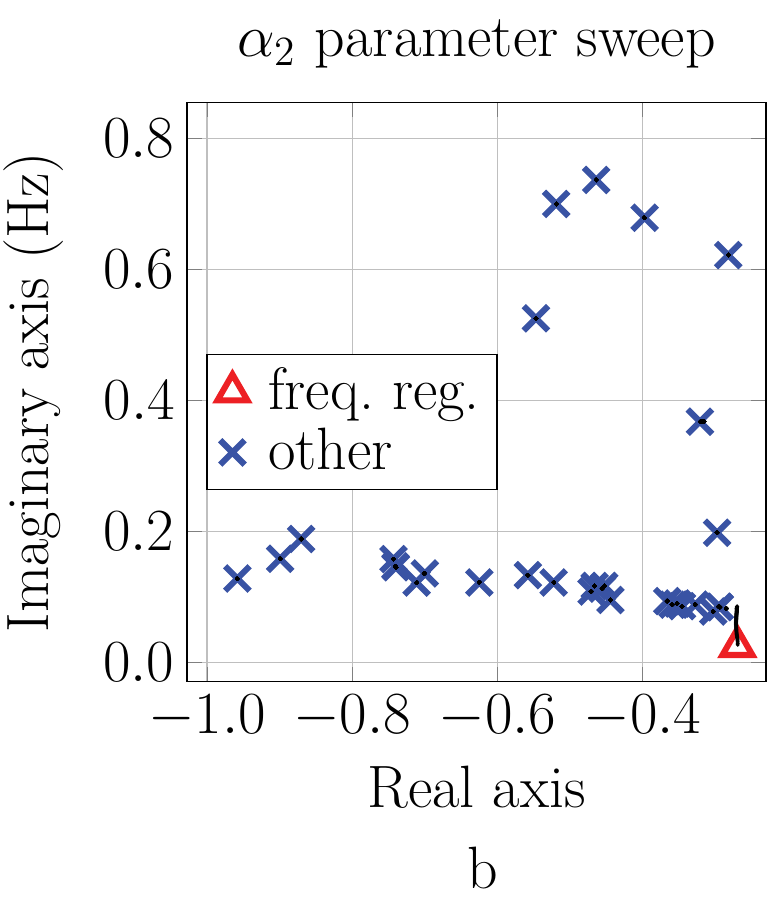}
        \label{fig:mini_alpha2_sensitivity}
    }
    \caption{Sensitivity of the system oscillatory modes to the tuning
        parameters. The frequency regulation mode, represented by the red
        triangle, is much more sensitive to changes in $\alpha_{2}$ than
        $\alpha_{1}$. The converse is true for the other modes. The shaded
        patch in (a) shows the axis range of
        (b).
    }
    \label{fig:mini_pole_sensitivity}
\end{figure}

Linear analysis allows us to develop criteria that are
\textit{necessary but not sufficient} for ensuring acceptable
transient stability control performance.  For example, theory dictates
that positive synchronizing torque drives the electromechanical modes
of oscillation upward in the complex plane, indicating an increase in
frequency.  Furthermore, the open-loop frequency response for each
actuator must obey the Nyquist stability criterion.
Fig.~\subref*{fig:mini_alpha1_sensitivity} shows how the eigenvalues
of the miniWECC respond when $\alpha_{1}$ is swept over the interval
$[0,1]$ uniformly for all actuators, where ${\alpha_{2}=0}$ and
${K=30}$.  The electromechanical modes that are responsive to the
control are either pushed directly upward, indicating pure
synchronizing torque, or up and to the left, indicating a combination
of damping and synchronizing.  The frequency regulation mode, marked
with a red triangle, is not sensitive to $\alpha_{1}$.  Conversely,
Fig.~\subref*{fig:mini_alpha2_sensitivity} shows that as $\alpha_{2}$
is swept over $[0,0.1]$, the frequency regulation mode moves upward
while the inter-area and local modes are unaffected.  The phase lead
introduced in the $\alpha_{2}$ path ensures that the angle of
departure of the frequency regulation mode is directly vertical, as
opposed to slightly to the right.  Recall from root locus analysis
that the angle of departure is the initial angle at which a
closed-loop pole travels in response to a change in the swept
parameter~\cite{stapleton:64}.

\begin{figure}[!t]
    \centering
    \includegraphics[width=0.631\textwidth]{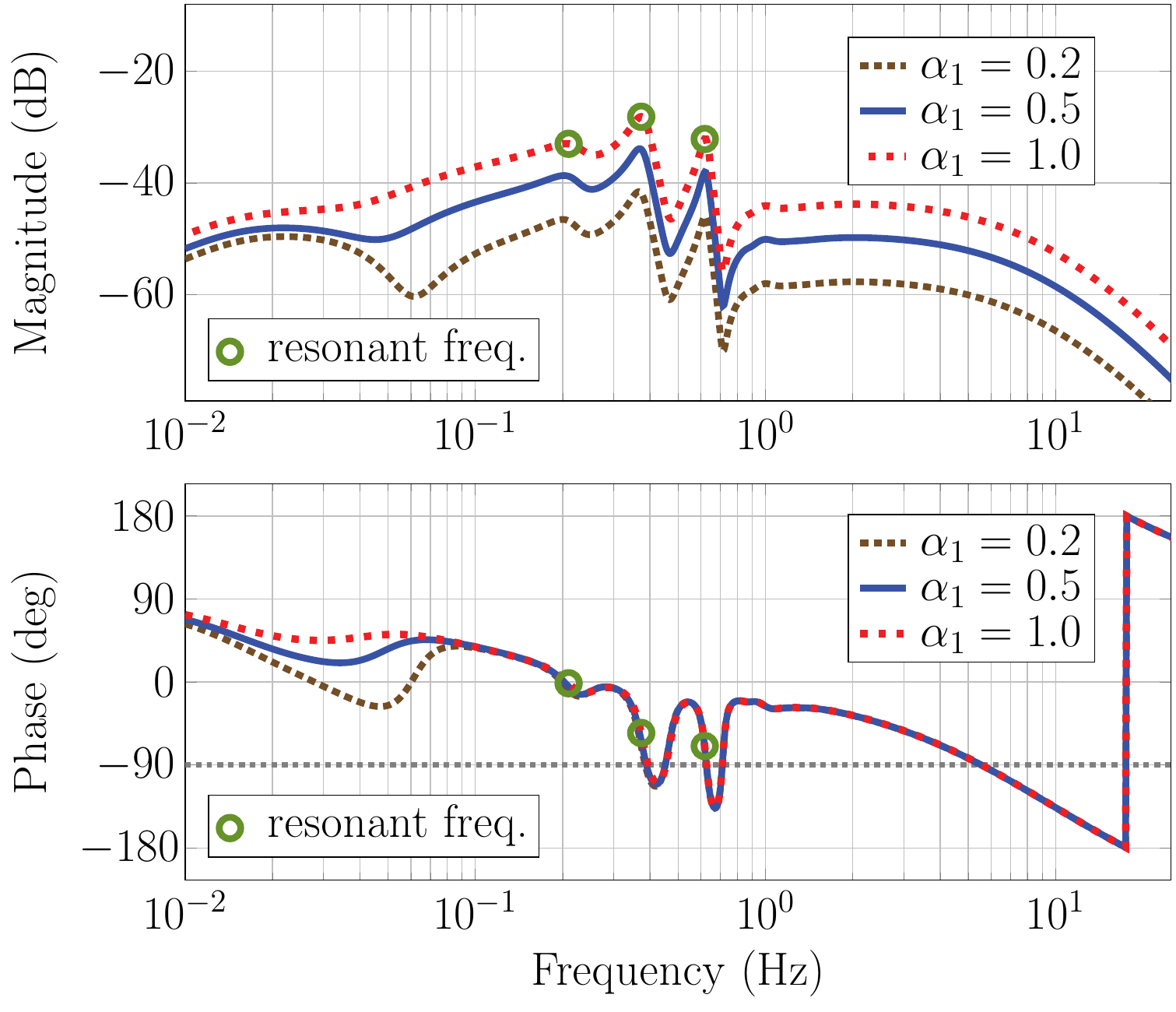}
    \caption{The open-loop frequency response between a change in
        $P_{\mathrm{ref}}$ and $p_{s}$ for the ESS located in British
        Columbia, where ${\alpha_{2} = 0.01}$ for all traces.  At a given
        frequency, the phase response indicates whether the controller
        provides damping torque (\SI{0}{\degree}), synchronizing torque
        (\SI{-90}{\degree}), or some combination thereof.
    }
    \label{fig:mini_alpha1_bode}
\end{figure}

\begin{figure}[!t]
    \centering
    \includegraphics[width=0.631\textwidth]{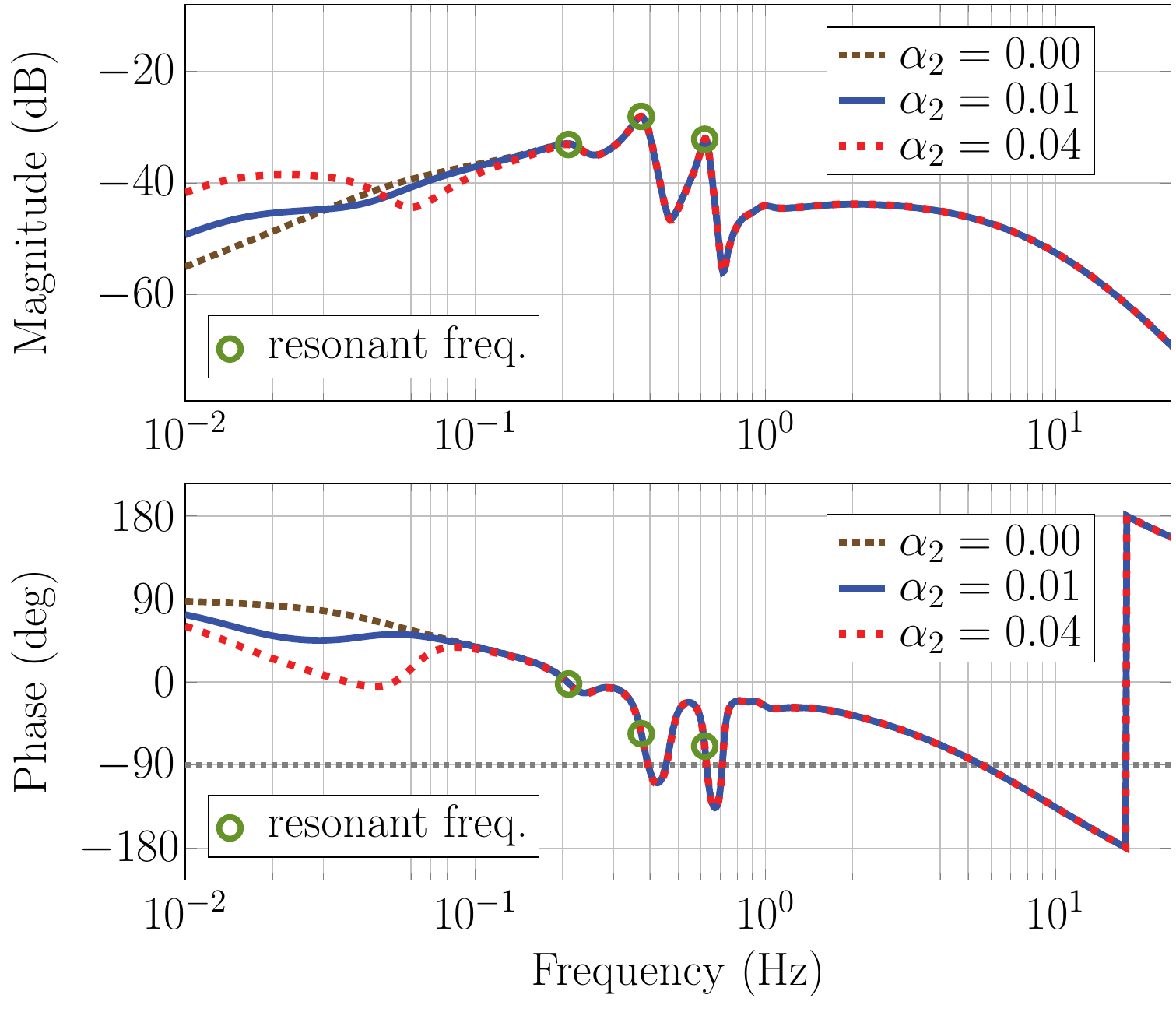}
    \caption{The open-loop frequency response between a change in
        $P_{\mathrm{ref}}$ and $p_{s}$ for the ESS located in British
        Columbia, where ${\alpha_{1} = 1.0}$ for all traces.  At a given
        frequency, the phase response indicates whether the controller
        provides damping torque (\SI{0}{\degree}), synchronizing torque
        (\SI{-90}{\degree}), or some combination thereof.
    }
    \label{fig:mini_alpha2_bode}
\end{figure}

\Cref{fig:mini_alpha1_bode} shows the open-loop frequency response
between a change in $P_{\mathrm{ref}}$ and the controller output
$p_{s}$ for the ESS located in British Columbia. The traces correspond
to various values of $\alpha_{1}$, where ${\alpha_{2}=0.01}$ is held
fixed.  If the controller were to provide pure synchronizing torque
across its entire bandwidth, then the phase response would transition
through \SI{-90}{\degree} at each resonant frequency.  We observe that
the North-South A mode just above \SI{0.20}{\hertz} is not strongly
observable in the amplitude response. At this frequency, the phase is
near \SI{0}{\degree} indicating that the controller provides more
damping torque than synchronizing.  At the other two highlighted
resonances, the North-South B mode near \SI{0.35}{\hertz} and the
British Columbia mode near \SI{0.63}{\hertz}, the phase response is
much closer to \SI{-90}{\degree}, indicating a stronger synchronizing
response.  Although~\cref{fig:mini_alpha1_bode} corresponds to an
individual actuator, these observations mirror the angles of departure
shown in~\cref{fig:mini_pole_sensitivity}.  Varying $\alpha_{1}$
serves to adjust the amplitude response over the range of frequencies
above the frequency regulation mode.  In contrast, varying
$\alpha_{2}$ adjusts the amplitude response in the neighborhood of the
frequency regulation mode.  \Cref{fig:mini_alpha2_bode} shows the
open-loop frequency response of the same ESS for various values of
$\alpha_{2}$, where ${\alpha_{1}=1.0}$ for all traces.  Because the
$\alpha_{2}$ deviations
${\widetilde{\theta}(t)-\widetilde{\theta}(t_{0})}$ may be
considerably larger than the $\alpha_{1}$ deviations
${\theta_{j}(t)-\widebar{\theta}_{j}(t)}$, the tuning parameters
should be set such that ${\alpha_{2}\ll\alpha_{1}}$.  In the analysis
presented in \cref{sec:nminus}, we set these parameters to
${\alpha_{1}=1.0}$ and ${\alpha_{2}=0.01}$ for all actuators.

While $\alpha_{1}$ and $\alpha_{2}$ are useful tools for shaping the
amplitude response, they have a limited impact on the phase.  To
address this limitation, a lead-lag compensation stage is included in
each branch of the control structure shown in
\cref{fig:main_control_structure}.  As a consequence of the Nyquist
stability criterion, a guiding principle when tuning the controller is
to keep the phase response within $\pm\text{\SI{90}{\degree}}$ from dc
through the position of the highest resonant frequency.  This ensures
that critical frequencies, where the phase response reaches
\SI{-180}{\degree}, will not coincide with any resonant peaks in the
amplitude response~\cite{pierre:19}.  A second principle is to aim for
a phase response that transitions through \SI{-90}{\degree} at the
resonant frequencies~\cite{dudgeon:07}. This ensures that the
controller primarily provides synchronizing torque, which it must if
it is to improve transient stability.  As discussed above, if the
phase response does not transition through \SI{-90}{\degree} at a
resonant peak, then it is desirable for it fall between
$\lbrack\text{\SI{-90}{\degree}},\text{\SI{0}{\degree}}\rbrack$ so
that it provides a combination of synchronizing and damping
torque~\cite{pourbeik:96}.

The broad hump in the amplitude response near \SI{0.02}{\hertz} shown
in Figs.~\ref{fig:mini_alpha1_bode} and \ref{fig:mini_alpha2_bode}
corresponds to the frequency regulation mode. Although it manifests
itself as a mild resonance, the tuning considerations are somewhat
different than for the inter-area modes.  Time- and frequency-domain
objectives must be given equal weight.  Values of $\alpha_{2}$ larger
than those shown in \cref{fig:mini_alpha2_bode} may cause the
controller to interfere with turbine governing and/or automatic
generation control (AGC) during transient disturbances. They may also
deprive the $\alpha_{1}$ component of the control error of the
modulation headroom necessary to keep individual generators in
synchronism with the center of inertia. Thus, as stated above, it is
advisable to ensure that $\alpha_{2}\ll\alpha_{1}$.  Rather than
tuning the control response in relation to the frequency regulation
mode exclusively using Bode diagrams, it may be preferable to
incorporate a combination of modal sensitivity analysis, as shown in
\cref{fig:mini_pole_sensitivity}, and time-domain simulation. The
compensation should be configured such that the angle of departure of
the frequency regulation mode is as close to vertical as possible as
$\alpha_{2}$ increases. This ensures that the controller provides
synchronizing torque at the modal frequency~\cite{pourbeik:96}.

The tuning considerations discussed here are applicable primarily to
large interconnections. In systems where an individual machine
makes up a substantial percentage of the overall system
inertia/capacity, different tunings may be necessary. In particular,
the relative value of $\alpha_{1}$ to $\alpha_{2}$ may be somewhat
different, with larger values of $\alpha_{2}$ potentially being viable
in some cases.

\clearpage
\section{$\text{N-1}$ Contingency Analysis}
\label{sec:nminus}
In this section, we conduct $\text{N-1}$ contingency analysis to study
the impact of the control strategy for a range of disturbances.  This
analysis serves as an example of the type of simulation protocol that
would be required to verify that a set of actuators, tuned in a
particular way, yields acceptable control performance at a given
operating point.  We simulated \num{28} generator trips, \num{80}
transmission line faults (three-phase bolted, cleared after
\num{6}~cycles), and \num{19} losses of load (\SI{50}{\percent} of
apparent power). Each contingency was simulated three times: in open
loop, closed loop, and in closed loop with nonideal communication.  We
employed a statistical communication model that accounts for delays,
jitter, and measurement noise.  The expected delay for each sensor was
uniformly distributed between \num{67} and \SI{250}{\milli\second}.
The high end of this range is more than double the maximum value
reported in a recent real-world experiment
(\num{69}--\SI{113}{\milli\second})~\cite{pierre:19}.

The actuators were placed and configured as
in~\cref{sec:time_domain_analysis} with a \SI{100}{\mega\watt} ESS
located at each of the \num{19} load centers.  The tuning parameters
and gain were set uniformly for all actuators such that
${\alpha_{1}=1.0}$, ${\alpha_{2}=0.01}$, and ${K=10}$.  These example
values were selected based on the time-domain sensitivity analysis
performed in \cref{sec:time_domain_analysis} and the frequency-domain
tuning considerations in \cref{sec:frequency_domain_analysis}.  As a
result of the device power rating, each actuator can inject (or
withdraw) a maximum of \SI{100}{\mega\watt}. During a severe
disturbance, actuators in different parts of the system respond
differently, with some injecting and some withdrawing power over the
first swing of the transient.

\begin{table}[!b]
    \renewcommand{\arraystretch}{1.15}
    \centering
    \caption{$\text{N-1}$ Contingency Analysis First Swing Summary\strut}
    \label{tab:nminus_summary}
    \begin{tabular}{l r r r}
        \toprule
        Event & Number of & Improvement & Mean decrease \\
              & first swings & rate (\si{\percent}) & in $\Delta\delta$ (\si{\percent}) \\
        \midrule
        Loss of load & \num{646} & \num{99.7} & \num{18.0} \\
        Gen.\ trip & \num{924} & \num{99.5} & \num{18.4} \\
        Fault/line clearing & \num{2720} & \num{99.3} & \num{16.8} \\
        \midrule
        Total & \num{4290} & \num{99.4} & \num{17.3} \\
        \bottomrule
    \end{tabular}
\end{table}

\Cref{tab:nminus_summary} provides a summary of the results. It
compares the open-loop case to the closed-loop case with nonideal
communication.  Here the \textit{first swing} of the generator
response is defined as the magnitude of the initial extremum of
$\Delta\delta_{i}$ following the disturbance.  The control strategy
reduced the magnitude of \SI{99.4}{\percent} of first swings by an
average of \SI{17.3}{\percent}.  For the very small fraction of first
swings (\SI{0.6}{\percent}) that did not decline, the average increase
was \SI{1.1}{\percent}.  The largest observed improvement in the
magnitude of the first swing was \SI{-21.37}{\degree}, while the
worst-case degradation was \SI{0.95}{\degree}.

\begin{figure}[!t]
    \centering
    \includegraphics[width=0.631\textwidth]{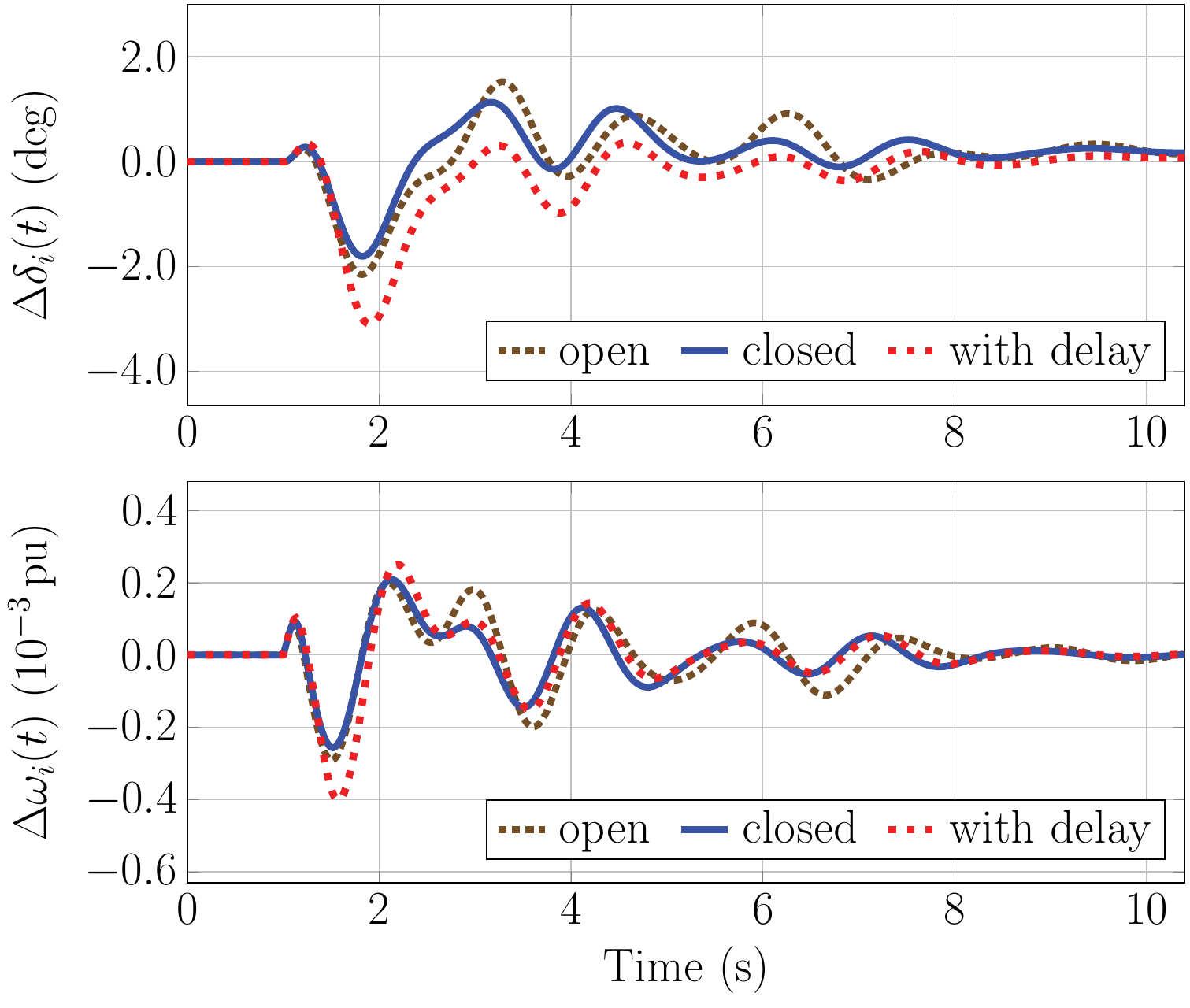}
    \caption{Simulated state trajectories for generator G31 in Wyoming
        in response to a \num{6}-cycle fault on the California-Oregon
        Intertie. This represents the worst-case degradation in the
        first swing observed in the $\text{N-1}$ contingency
        analysis.
    }
    \label{fig:mini_worst_first_swing}
\end{figure}

The worst-case response was observed at generator G31 in Wyoming in
response to a fault on the California-Oregon Intertie.  In the case
with nonideal communication, this angle response represented the worst
first-swing degradation both in absolute terms, as measured in
degrees, and on a percentage basis.  \cref{fig:mini_worst_first_swing}
shows the angle and speed trajectories for G31 with respect to the
center of inertia in response to the fault.  In the closed-loop case
with ideal communication, the generator response is effectively the
same as in the open-loop case; however, in the case with nonideal
communication, it is degraded slightly. The scale of the plots is
zoomed in to show the differences between the observed trajectories.
The magnitude of the first-swing angle response increased by
\SI{0.95}{\degree}, and the speed response by \SI{1e-4}{\pu}.  This
generator is far away from the fault and not at serious risk of
instability for this disturbance.

In these examples, the IBRs are located at the major load centers, as
shown in \cref{fig:miniwecc_map}.  Since much of the power generated
in Montana and Wyoming is exported elsewhere, there are no IBRs
modeled in either of those states.  The response of generator G31
could be improved by placing a control actuator near it.  This
analysis indicates that even in the case where a generator is far away
(electrically) from the IBRs, it frequently experiences a moderate
improvement in its first-swing transient response as a result of the
control. In cases where the response does not improve, it does not
appear to be significantly degraded.

\subsection{In-Depth Fault Analysis}
\begin{figure}[!t]
    \centering
    \includegraphics[width=0.631\textwidth]{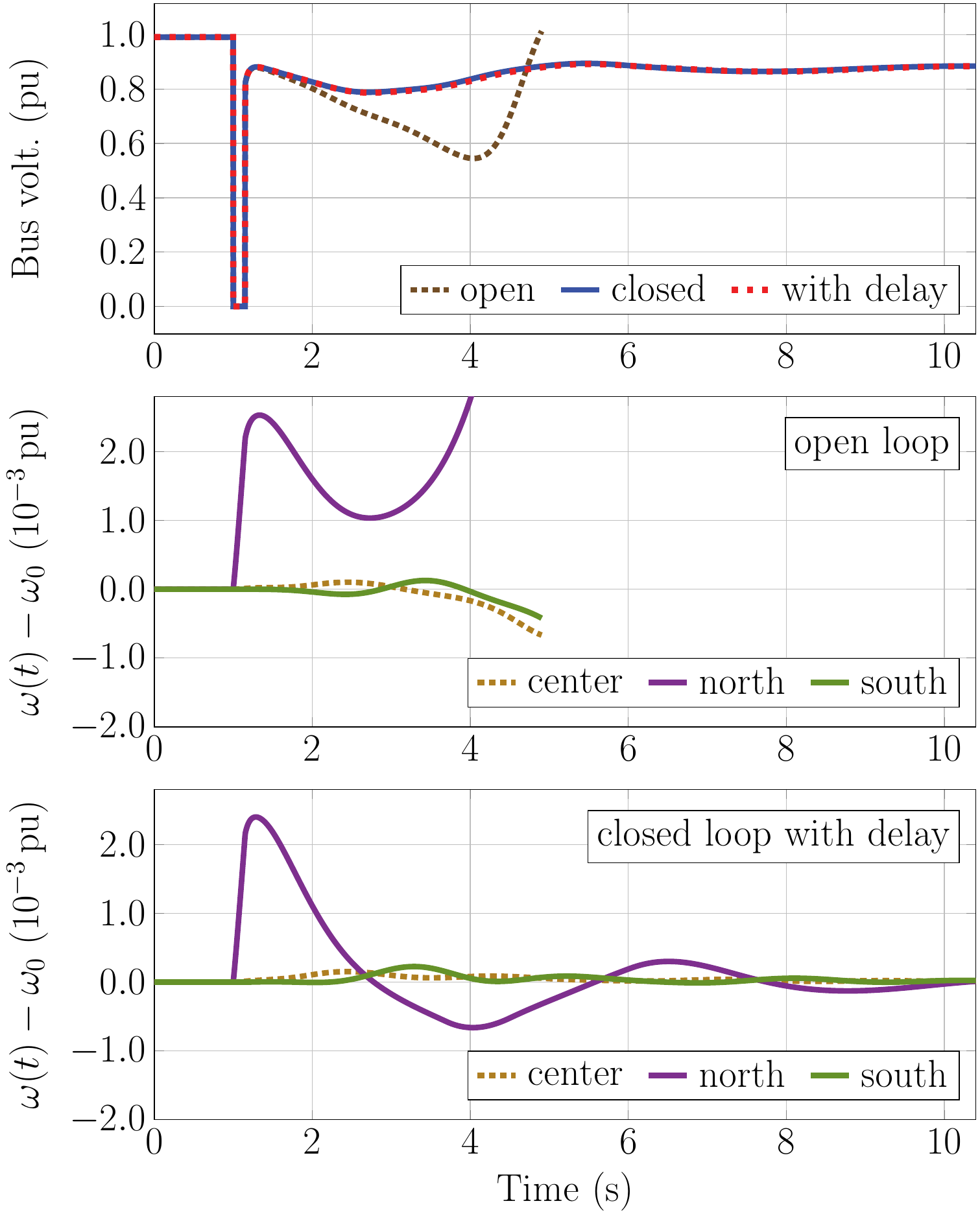}
    \caption{Time-domain simulations of a \num{9}-cycle fault near
	generator G34 in Alberta. The top subplot shows the voltage magnitude
	at the faulted bus. The bottom two subplots show the LTI speed
        deviations of G34 in Alberta (north) and G23 in San Diego (south)
        compared with the center of inertia.
    }
    \label{fig:mini_alberta_fault_time}
\end{figure}

To provide analysis of a representative contingency, we study a
\num{9}-cycle fault near generator G34 in Alberta.  For this fault,
the control strategy improves the critical clearing time from \num{6}
cycles to \num{10} cycles.  \Cref{fig:mini_alberta_fault_time} plots
the state response in the time-domain, where the upper subplot shows
the voltage magnitude at the faulted bus. The lower subplots show the
LTI speed deviations (i.e., ${\omega(t)-\omega_{0}}$) for G34 in
Alberta, G23 in San Diego, and the center of inertia.  For the
closed-loop cases, the results with delay effectively overlap those
without.  In open loop, the generator in Alberta loses synchronism.
When this occurs, G23 in San Diego begins oscillating against the
center of inertia.  In practice, out-of-step or overspeed protection
would likely trip the critical unit in Alberta before it lost
synchronism. This protection may mitigate the oscillations observed in
San Diego. The bottom subplot shows the speed deviations in closed
loop, including the effect of delays and noise.  In this case, the
generator in Alberta remains synchronized and the oscillations in San
Diego subside.

\begin{figure}[!t]
    \captionsetup[subfloat]{farskip=0pt}
    \centering
    \subfloat{
        \centering
        \includegraphics[width=0.385\textwidth]{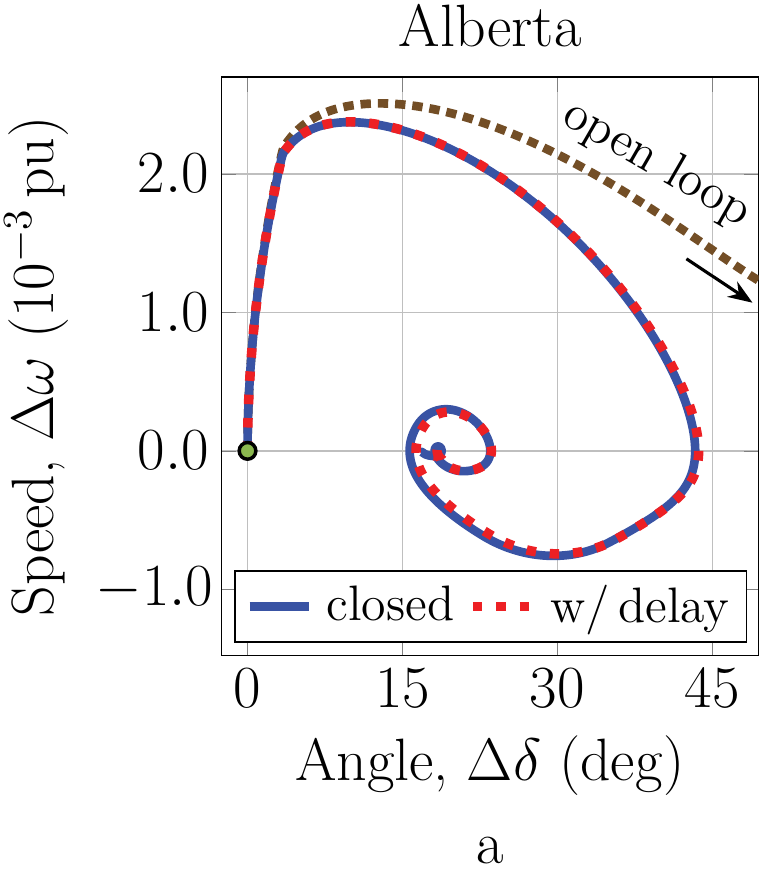}
        \label{fig:mini_north_phase_plane}
    }\qquad
    \subfloat{
        \centering
        \includegraphics[width=0.385\textwidth]{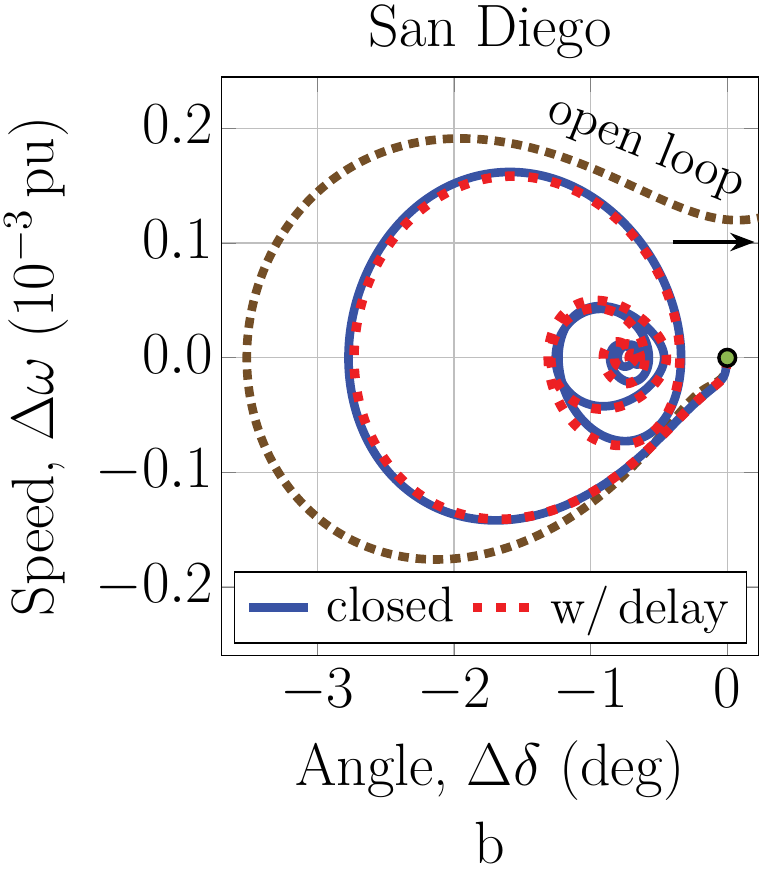}
        \label{fig:mini_south_phase_plane}
    }
    \caption{Phase plane analysis in the center-of-inertia reference
        frame for a \num{9}-cycle fault near generator G34 in
        Alberta. Subfigure (a) shows the behavior of G34 in
        Alberta, and (b) G23 in San Diego.  All curves begin
        at the origin.
    }
    \label{fig:mini_phase_planes}
\end{figure}

We can gain additional insight into this fault by studying the system
response in the angle domain.  The phase portraits shown in
\cref{fig:mini_phase_planes} plot the LTV speed deviations
$\Delta\omega_{i}$ versus $\Delta\delta_{i}$ for the two machines
discussed above.  In open loop, the curves do not arrive at a
post-disturbance equilibrium; however, in closed loop they do. This
indicates that the control action expands the region of attraction to
encompass the point in the plane where each generator resides
immediately after the fault.

\begin{figure}[!t]
    \centering
    \includegraphics[width=0.631\textwidth]{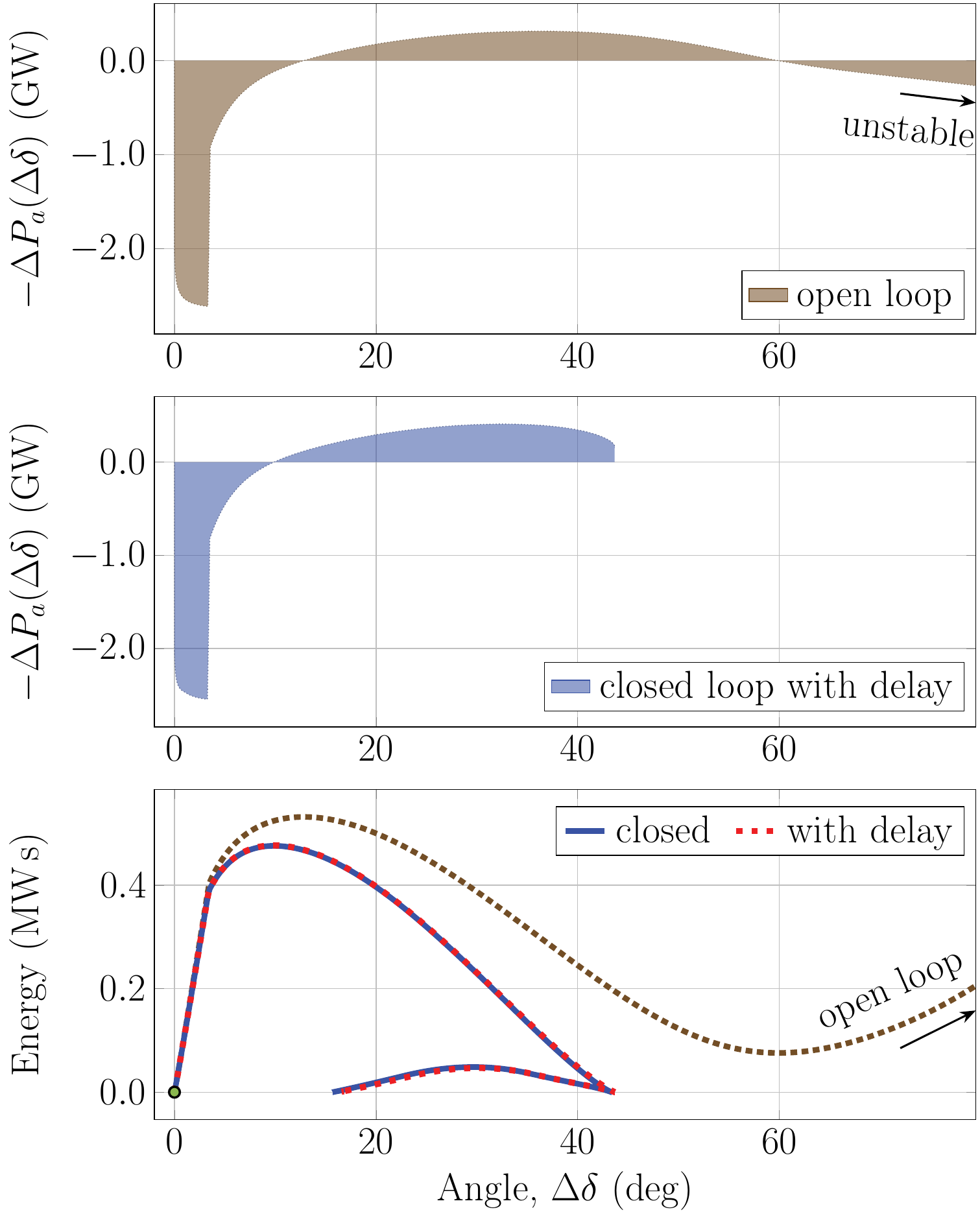}
    \caption{Accelerating power analysis in the center-of-inertia
        reference frame for a \num{9}-cycle fault near generator G34 in
        Alberta. The top subplot shows the accelerating and decelerating areas
        in open and closed loop. The bottom subplot shows the integral of
        accelerating power as a function of $\Delta\delta$.
    }
    \label{fig:mini_alberta_equal_area}%
\end{figure}

Using this perspective, we can also compare the accelerating areas.
Under the classical model, the accelerating power of the $i$th machine
in the center-of-inertia reference frame is
\begin{equation}
    \label{eq:coi_accelerating_power}
    \Delta P_{a}^{i}(t) = P_{m}^{i}(t)-P_{e}^{i}(t)-\frac{H_{i}}{H_{\scriptscriptstyle{T}}}
    \Biggl\lbrack \sum_{k\in\mathcal{K}}P_{m}^{k}(t)-P_{e}^{k}(t) \Biggr\rbrack,
\end{equation}
where $\mathcal{K}$ is the set of all online synchronous
machines~\cite{michel:83}.  The top subplot
of~\cref{fig:mini_alberta_equal_area} shows the accelerating power of
G34 in Alberta as a function of $\Delta\delta_{i}$, and the bottom the
integral of $\Delta P_{a}^{i}$ over $\Delta\delta_{i}$.  Assuming the
damping constants are negligible, this integral yields the kinetic
energy in the center-of-inertia reference frame
$H_{i}\Delta\omega_{i}^{2}$.  In open loop, the decelerating area
is insufficient to cancel the accelerating area, and the machine pulls
away from the stable equilibrium. In the bottom subplot, this
manifests itself as the failure of the kinetic energy curve to reach
the $x$-axis, i.e., zero energy.  The control action reduces the
accelerating area and expands the decelerating area, as shown in
blue. This allows the kinetic energy curve to reach the $x$-axis,
where the machine begins its second swing.

\clearpage
\section{Simplified Path Rating Study}
\label{sec:coi_rating}

In practice, the type of contingency analysis presented in
\cref{sec:nminus} plays an important role in establishing ratings for
stability-limited transmission corridors. One such corridor in the
Western Interconnection is the California-Oregon Intertie.  By
choosing a plausible critical contingency for this corridor, we can
estimate the impact of the control strategy on the transfer
limit. With north-to-south flows, we will assume the critical
contingency is a trip of G18 in the miniWECC, a large gas-fired
generator near the southern end of the intertie.  At the initial
operating point G18 supplies \SI{5}{\giga\watt} of real power, or
roughly \SI{4.5}{\percent} of the system load.  Thus, the interface
flows must change substantially in order for the system to remain
stable following the disturbance.  In cases where the system
stabilized, the flow on the California-Oregon Intertie was observed to
increase by more than \SI{1}{\giga\watt} following the disturbance.
This is expected because the loss of G18 results in a generation
deficit in California.  When the power transfer on the intertie
exceeds its maximum safe level, the system lacks the stiffness to
accommodate the sudden increase in flows caused by the generation
trip. The voltage magnitudes at the ends of the intertie fall
precipitously, and the power transfer never stabilizes.  When this
occurs, the system begins to separate with the Pacific Northwest and
the northern half of the system pulling away from California and the
southern half.

\begin{figure}[!t]
    \centering
    \includegraphics[width=0.631\textwidth]{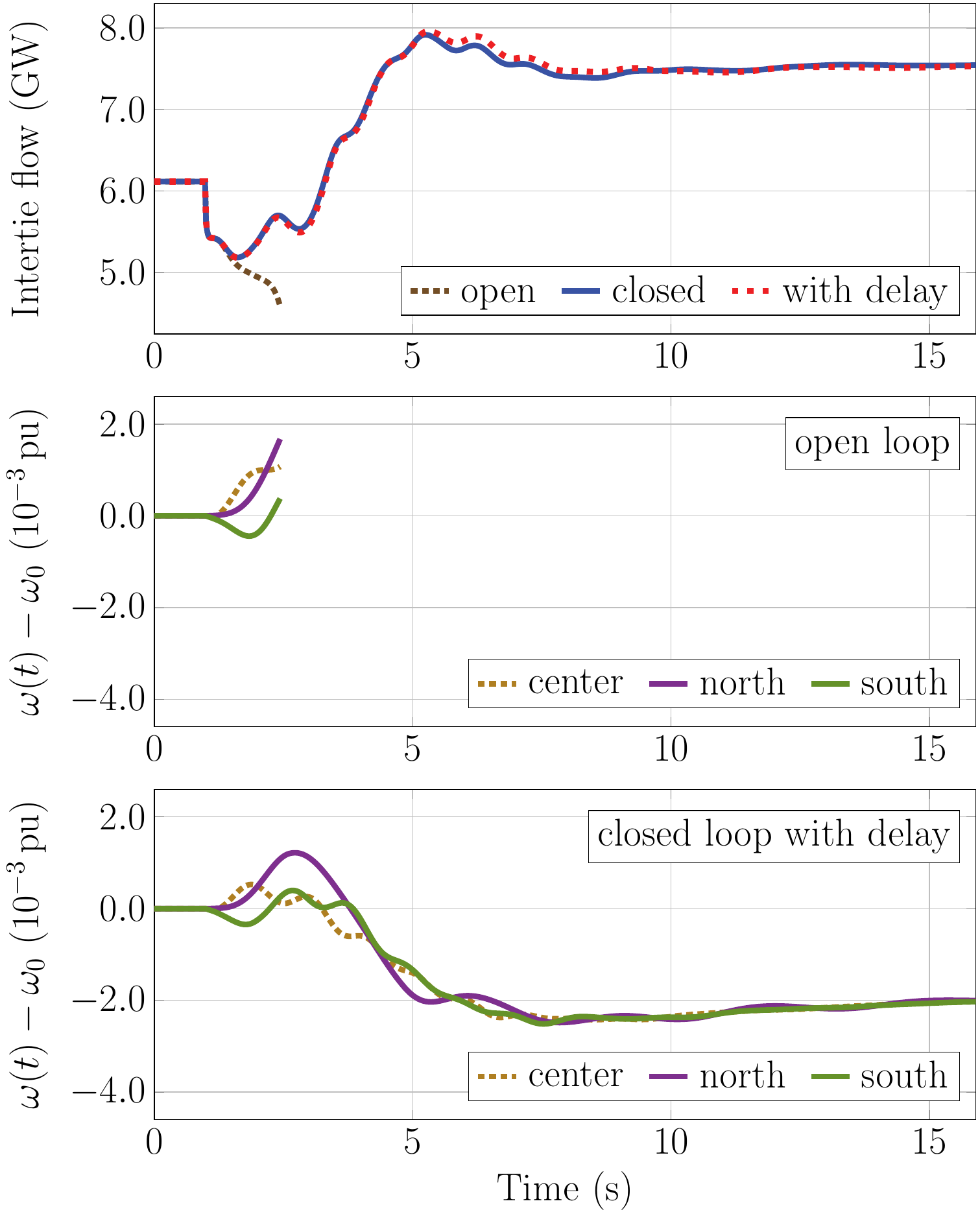}
    \caption{Time-domain simulations for a trip of generator G18 near
        the southern end of the California-Oregon Intertie.  The upper subplot
        shows the active power flow on the intertie from north to south.  The
        bottom two subplots show the LTI speed deviations of G34 in Alberta
        (north) and G23 in San Diego (south) compared with the center of
        inertia.
    }
    \label{fig:mini_coi_rating_time}
\end{figure}%

For this contingency, we find that the maximum transfer that yields a
stable response in the open-loop case is \SI{5.3}{\giga\watt}.  To
determine this limit, we incrementally increased the loading in
California and redispatched the generation in the Pacific Northwest.
In the closed-loop case, the controllers produce supplemental
synchronizing torque by reducing deviations in the power angle across
the intertie.  Immediately following the generation trip, the energy
storage systems in California discharge, helping to offset the loss of
G18.  The storage systems in the Pacific Northwest and Canada charge,
which increases the load in the northern half of the system in an
effort to reduce north-to-south intertie flows.  In closed loop, the
transfer limit increases from \SI{5.3}{\giga\watt}
to~\SI{6.1}{\giga\watt}.  Hence, in this example, \SI{1.9}{\giga\watt}
of energy storage is able to increase the transfer limit by
roughly~\SI{800}{\mega\watt}. In reality, this result would need to be
confirmed for various operating conditions and multiple critical
contingencies.  The intent of this analysis is not to identify the
path rating for a particular operating condition, but rather to gauge
the magnitude and sign of the change in the path rating as a result of
control.  \Cref{fig:mini_coi_rating_time} shows time-domain simulation
results for a trip of G18 when the steady-state active power transfer
on the intertie is \SI{6.1}{\giga\watt}.  As explained above, in the
open-loop case the power transfer never stabilizes after the
disturbance. When the system goes unstable the power flow fails to
converge, halting the simulation.  In the closed-loop case, the
control action allows the system to safely reach a post-disturbance
equilibrium.

\clearpage
\section{Conclusion}
\label{sec:conclusion}

In this paper, we developed and demonstrated a control strategy that
modulates the active power injected by inverter-based resources.  For
testing, we employed a reduced-order dynamic model of the Western
Interconnection.  The key points can be summarized as follows:

\begin{itemize}
    \item The proposed method was based on a time-varying
    representation of the equations of motion for a synchronous machine.
    \item The control strategy uses IBRs to inject active power into the
    system that is in phase with the LTV angle deviations.
    \item For each IBR, the injection drives the bus voltage angle
    toward a desired trajectory that tracks the angle of the center of inertia.
    \item The feedback signals are constructed using bus voltage phasor data
    collected from wide-area measurement systems.
    \item While generator data may be used if it is available, it is not
    required to produce an estimate of the center-of-inertia states
    that is sufficiently accurate for control.
    \item This control strategy improves system reliability by
    helping to keep each generator in synchronism with the center of
    inertia.
    \item In addition, this approach can increase capacity utilization
    on stability-limited transmission corridors, allowing existing
    infrastructure to be used more efficiently.
\end{itemize}
We are currently conducting ongoing research examining the performance
of the developed control strategy in relation to alternative
technologies/approaches.  This work is also exploring techniques and
considerations for setpoint management, i.e., when and how to update
the constants in~\cref{eq:meas_angle_trajectory} as the dispatch
pattern changes.  Future work will investigate improved methods for
characterizing transient stability margins in nonlinear systems.

%

\clearpage
\bibliographystyle{ieeetr}
\bibliography{./bib/ltv_transient_revision}

\end{document}